\documentclass[10pt,superscriptaddress,nofootinbib,twocolumn,aps,pra]{revtex4-2}

\usepackage{amsthm,amsmath,amssymb}
\usepackage{mathrsfs}
\usepackage{appendix}
\usepackage{amstext}
\usepackage{graphicx}
\usepackage{url}
\usepackage{float}
\usepackage{bm}
\usepackage[usenames,dvipsnames]{color}
\newcommand{\bra}[1]{\mbox{$\left\langle #1 \right|$}}
\newcommand{\ket}[1]{\mbox{$\left| #1 \right\rangle$}}
\usepackage[colorlinks=true,citecolor=blue,urlcolor=black]{hyperref}
\usepackage{dcolumn}
\usepackage{natbib}
\usepackage{booktabs}

\begin{document}

\title{Homodyne Detection Quadrature Phase Shift Keying Continuous-Variable Quantum Key Distribution with High Excess Noise Tolerance}

\author{Wen-Bo Liu}
\author{Chen-Long Li}
\author{Yuan-Mei Xie}
\author{Chen-Xun Weng}
\author{Jie Gu}
\author{Xiao-Yu Cao}
\author{Yu-Shuo Lu}
\author{Bing-Hong Li}
\author{Hua-Lei Yin}\email{hlyin@nju.edu.cn}
\author{Zeng-Bing Chen}\email{zbchen@nju.edu.cn}
\affiliation{National Laboratory of Solid State Microstructures, School of Physics and Collaborative Innovation Center of Advanced Microstructures, Nanjing University, Nanjing 210093, China}

\begin{abstract}
Discrete-modulated continuous-variable quantum key distribution with homodyne detection is widely recognized for its ease of implementation, efficiency with respect to error correction, and its compatibility with modern optical communication devices.
However, recent studies report that the application of homodyne detection obtains poor tolerance to excess noise and insufficient transmission distance, hence seriously restricting the large-scale deployment of quantum secure communication networks.
In this paper, we propose a homodyne detection protocol using the quadrature phase shift keying technique.
By limiting information leakage, our proposed protocol enhances excess noise tolerance to a high level.
Furthermore, we demonstrate that homodyne detection performs better than heterodyne detection in quaternary-modulated continuous-variable quantum key distribution under the untrusted detector noise scenario.
The security is analyzed using the tight numerical method against collective attacks in the asymptotic regime.
Our results imply that the current protocol is able to distribute keys in nearly intercity area and thus paves the way for  constructing low-cost quantum secure communication networks.
\end{abstract}

\maketitle
\section{INTRODUCTION}
In recent years, the rapid development of quantum computing~\cite{arute2019quantum,zhong2020quantum} has signaled the coming quantum era, which threatens modern secure communications. For example, the most widely used public key cryptography, the Rivest-Shamir-Adleman cryptosystem~\cite{rivest1978method}, whose security relies on the complexity of factorizing a large number, can be cracked by the Shor algorithm~\cite{shor1994algorithms} with quantum computers. Fortunately, the one-time pad algorithm~\cite{shannon1949communication} can ensure information-theoretically secure communication provided that two remote users share a large number of identical secret keys. Quantum key distribution (QKD)~\cite{bennett1984quantum,ekert1991quantum} is the only solution that enables two distant users to share secure keys against the most general attacks~\cite{RevModPhys.84.621,Pirandola2020advances, RevModPhys.92.025002}.

Since the first protocol was proposed by Bennett and Brassard in 1984~\cite{bennett1984quantum}, a large number of discrete-variable quantum key distribution (DV-QKD) protocols, based on various discrete degrees of freedom of a single photon, have been developed, such as polarization~\cite{PhysRevLett.96.070502,comandar2016quantum,wei2020high} and time bin~\cite{yin2016measurement,boaron2018secure,yin2020experimental}. Many attempts have been made to realize the network deployment of DV-QKD~\cite{Pirandola2020advances, RevModPhys.92.025002,Joshieaba0959,tang2016measurement}. Compared with DV-QKD, continuous-variable quantum key distribution (CV-QKD)~\cite{RevModPhys.84.621,diamanti2015distributing} has the competitive advantages of lower-cost implementations and higher secret key rates over the metropolitan areas, and it has thus attracted widespread attention.

CV-QKD encodes keys on the quadratures of the quantized electromagnetic field by preparing and measuring coherent states~\cite{grosshans2002continuous,grosshans2003quantum,yin2019phase} or squeezed states~\cite{cerf2001quantum}. From an experimental perspective, CV-QKD based on Gaussian modulation has made remarkable achievements~\cite{lodewyck2007quantum,qi2007experimental,fossier2009field, jouguet2013experimental,huang2016long,PhysRevA.102.032625,PhysRevX.5.041009, PhysRevX.5.041010,huang2015high,pirandola2015high,zhang2019integrated,zhang2020long}, such as the provision of security against all detection loopholes~\cite{pirandola2015high}, integration of all optical components on chips~\cite{zhang2019integrated}, and record-breaking distance up to 200 km~\cite{zhang2020long}. Because Gaussian modulation, which samples quadratures from Gaussian distribution, possesses $U(n)$ symmetry, it has a relatively complete security analysis~\cite{silberhorn2002continuous,garcia2009continuous,leverrier2010finite, leverrier2015composable,PhysRevResearch.3.013279} based on the
optimality of Gaussian attacks~\cite{PhysRevLett.97.190502,PhysRevLett.97.190503}. However, the continuous modulation of quadratures is not experimentally feasible ~\cite{lupo2020towards}. Practical implementations sample from a discrete and finite distribution of coherent states to approximate the Gaussian distribution, resulting in an unclosed gap between security analysis and experimental implementation. Moreover, Gaussian modulation requires complex post-processing procedures mainly because of the low signal-to-noise ratio and complex data structure, which often consumes large amounts of computing resources~\cite{jouguet2013experimental,PhysRevApplied.12.054013}.

Unlike Gaussian modulation, discrete modulation~\cite{leverrier2009unconditional,xuan200924,hirano2017implementation,lin2020trusted, ghalaii2020discrete} can sample from a quite small set of coherent states with different phases and, thus, can perform an easier error correction. The security proof of discrete modulation carefully considers the effect of discrete distribution of prepared states. Despite that discrete modulation lacks $U(n)$ symmetry, several effective numerical~\cite{ghorai2019asymptotic,lin2019asymptotic,hu2021robust} and analytical methods~\cite{PhysRevA.103.012412,denys2021explicit} have emerged for security analysis, and robustness to collective attacks in the asymptotic regime has been verified. Recently, the security proof of a binary-modulated protocol has been proposed in the finite-key-size regime against general coherent attacks~\cite{matsuura2021finite}. With security improvement and simplified implementations, discrete modulation becomes a trend of CV-QKD~\cite{Pirandola2020advances, RevModPhys.92.025002}.

Currently, for CV-QKD implementations, homodyne detection is a relatively mature technology with the advantages of simple structures, stable performance, high efficiency, and low noise, making it widely adopted~\cite{lodewyck2007quantum,qi2007experimental,fossier2009field, jouguet2013experimental,huang2016long,PhysRevA.102.032625,PhysRevX.5.041009, PhysRevX.5.041010,huang2015high,pirandola2015high,zhang2019integrated,zhang2020long}. However, the existing homodyne detection protocol for discrete modulation, specifically quaternary modulation~\cite{lin2019asymptotic}, performs poorly in terms of excess noise tolerance, which is far from practical demands.

In this paper, we present a quaternary-modulated protocol with homodyne detection. Unlike in previous quaternary-modulated protocols~\cite{ghorai2019asymptotic,lin2019asymptotic}, we prepare each signal with the phase chosen from $\{\frac{\pi}{4}, \frac{3\pi}{4}, \frac{5\pi}{4}, \frac{7\pi}{4}\}$, which can be implemented using classical quadrature phase shift keying (QPSK) technology. We require the sender to generate raw keys according to quadrature choices of the receiver. According to this interesting operation, our protocol shows good tolerance to excess noise and promises significantly longer transmission distances than the previous homodyne detection protocol~\cite{lin2019asymptotic}. In addition, the secret key rate of the proposed protocol is comparable to that of the well-performed heterodyne detection protocol~\cite{lin2019asymptotic}. Considering the imperfections of detectors, our protocol even performs better than the heterodyne detection protocol.

We introduce our protocol in Section~\ref{sec:pro}, with an instruction of a possible experimental setup. In Section~\ref{sec:sec}, the formula of the secret key rate is given against collective attacks in the asymptotic regime. In Section~\ref{sec:sim}, we numerically simulate the performance of our proposed protocol, which demonstrates an increase in transmission distance and secret key rate. Further discussions are presented in Section~\ref{sec:dis}.

\section{PROTOCOL DESCRIPTION}\label{sec:pro}
\begin{figure}[t]
\centering
\includegraphics[width=8.6cm]{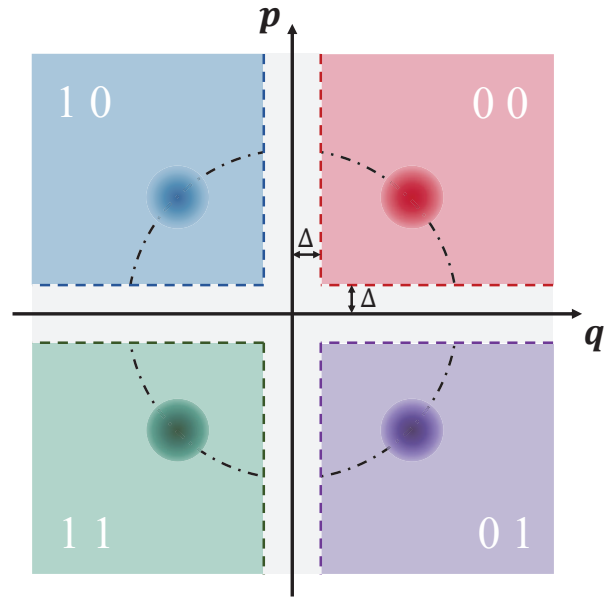}
\caption{Quaternary phase shift keying and key mapping of our protocol.
Alice randomly sends one coherent state with amplitude $\alpha$ and phase $\theta\in\{\frac{\pi}{4}, \frac{3\pi}{4}, \frac{5\pi}{4}, \frac{7\pi}{4}\}$ for each round, representing labels $\{00, 10, 11, 01\}$, respectively.
For each round, Bob randomly measures one quadrature from $\{\hat{q}, \hat{p}\}$.
He maps the outcome greater than $\Delta$ into bit 0 and the outcome smaller than $-\Delta$ into bit 1.
After the announcement, Alice records the first bit of the label as her key bit if a state is measured in $\hat{q}$ or the second bit of the label if a state is measured in $\hat{p}$.}\label{fig1}
\end{figure}

\begin{figure}[t]
\includegraphics[width=8.6cm]{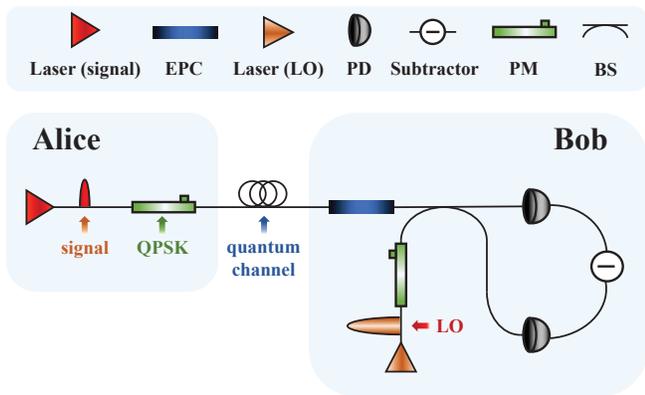}
\centering
\caption{Experimental schematic of our protocol. To prevent Eve from manipulating the local oscillator (LO) transmitted from Alice, we adopt the local LO scheme~\cite{huang2015high}. Abbreviations of components in the figure: EPC, electrical polarization controller; PD, photodetector; PM, phase modulator; QPSK, quadrature phase shift keying; BS, beam splitter.}\label{fig2}
\end{figure}

We illustrate our protocol in this section using the preparing and mapping method in Fig.~\ref{fig1} and the experimental setup in Fig.~\ref{fig2}.

\hangafter=1
\hangindent 1.5em
\noindent
\emph{(1) Preparation.}---For each round, the sender Alice prepares randomly one of four coherent states $\left\{\ket{\alpha e^{i\frac{\pi}{4}}}, \ket{\alpha e^{i\frac{3\pi}{4}}}, \ket{\alpha e^{i\frac{5\pi}{4}}}, \ket{\alpha e^{i\frac{7\pi}{4}}}\right\}$ with equal probabilities. She then sends the prepared state to the receiver Bob.

\hangafter=1
\hangindent 1.5em
\noindent
\emph{(2) Measurement.}---After receiving the state, Bob selects randomly a quadrature from $\{\hat{q},\hat{p}\}$ with equal probabilities and performs corresponding homodyne detection to obtain the measurement outcome.

\hangafter=1
\hangindent 1.5em
\noindent
\emph{(3) Announcement and parameter estimation.}---After sufficient rounds of the above two steps, Alice and Bob communicate through an authenticated public channel.
First, Bob announces his choice of quadrature for each round. Second, Alice and Bob randomly select a test subset from all rounds. For each round in the test subset, Alice discloses the state that she sends, and Bob discloses the outcome. Based on the information that they expose, they calculate the secret key rate under reverse reconciliation. If calculations show that no secret keys can be generated, they abort the protocol. Otherwise, they proceed.

\hangafter=1
\hangindent 1.5em
\noindent
\emph{(4) Raw key generation.}---Alice and Bob can obtain their raw keys using the remaining undisclosed rounds. Assuming that $M$ rounds remain, they number these rounds according to the order of sending. For the $k$th round, Alice labels the corresponding state $\ket{\phi_k}\in\left\{\ket{\alpha e^{i\frac{\pi}{4}}}, \ket{\alpha e^{i\frac{3\pi}{4}}}, \ket{\alpha e^{i\frac{5\pi}{4}}}, \ket{\alpha e^{i\frac{7\pi}{4}}}\right\}$ as $a_k\in\{00, 10, 11, 01\}$. Then, the raw key bit $x_k$ is equal to the first bit of the label $a_k$ when the round is measured in $\hat{q}$, and it is equal to the second bit of $a_k$ when the round is measured in $\hat{p}$. Thus, she obtains her string $\textbf{X}'=(x_1,..,x_k,...,x_M)$. For the $k$th round, Bob maps the outcome greater than $\Delta$ into the raw key bit $z_k=0$ and the outcome smaller than $-\Delta$ into $z_k=1$. The outcome with other values is mapped into $z_k=\bot$. Thus, he obtains his string $\textbf{Z}'=(z_1,..,z_k,...,z_M)$. $\Delta$ is a non-negative parameter and related to post-selection. A protocol without post-selection can set $\Delta=0$. If $\Delta > 0$, Bob communicates with Alice about positions of bits with value $\bot$ and they get rid of these positions from their strings $\textbf{X}'$ and $\textbf{Z}'$. Finally, they obtain their raw key strings $\textbf{X}$ and $\textbf{Z}$.

\hangafter=1
\hangindent 1.5em
\noindent
\emph{(5) Error correction and privacy amplification.}---Alice and Bob choose suitable methods to conduct error correction and privacy amplification. In the end, they generate secret keys.

A realistic experimental setup of our protocol is proposed in Fig.~\ref{fig2}. The required components are displayed with unnecessary details omitted. To prevent security loopholes of the LO, we adopt the local LO scheme with secure phase compensation~\cite{PhysRevX.5.041009, PhysRevX.5.041010, huang2015high}. Alice randomly~\cite{huang2015high} sends reference pulses and signal pulses. Signal pulses are modulated using QPSK technology, and are used to generate keys. Setting the phase of signal pulses before QPSK as the reference phase, Alice prepares reference pulses that are brighter than signal pulses and contribute nothing to the generation of keys. Reference and signal pulses are transmitted to Bob's side through quantum channels. The polarization of any pulse received by Bob is corrected by the electrical polarization controller.
Bob generates a local LO with the phase randomly chosen from $\{0,\frac{\pi}{2}\}$ using the phase modulator, which will be sent to the beam splitter together with signal pulses or reference pulses for interference. The photodetector with a subtractor reveals the interference outcomes, according to which Bob evaluates the exact phase difference between the reference phase and local LO. Then, Bob uses this phase information to conduct phase compensation~\cite{huang2015high}. Thereafter, Alice and Bob conduct post-processing described by steps \textit{(3)-(5)}, and we do not show this part in the figure.

\section{SECRET KEY RATE}\label{sec:sec}
To evaluate the secret key rate, we analyze the security of an equivalent entanglement-based protocol of our prepare-and-measure protocol. Then, we calculate the secret key rate against collective attacks in the asymptotic regime by applying the numerical method~\cite{lin2019asymptotic,hu2021robust}.

\subsection{Entanglement-based protocol}
An equivalent entanglement-based protocol of our prepare-and-measure protocol can be described below. Alice prepares an entangled state
\begin{equation}
\ket{\Phi}_{AA'}=\sum_{x}\sqrt{p_x}\ket{x}_A \ket{\phi_x}_{A'},
\end{equation}
where the subscripts $A$ and $A'$ represent two entangled systems. $x$ represents one of four labels in $\{00, 10, 11, 01\}$, and $p_x=0.25$ refers to the probability that the entangled state will collapse into the state labeled by $x$. $\{\ket{x}\}$ is the set of orthogonal bases in system $A$, which can be measured by a set of positive operator valued measurement $\{M^x = \ket{x}\bra{x}\}$. Four states in $\{\ket{\phi_x}\}$ are $\left\{\ket{\alpha e^{i\frac{\pi}{4}}}, \ket{\alpha e^{i\frac{3\pi}{4}}}, \ket{\alpha e^{i\frac{5\pi}{4}}}, \ket{\alpha e^{i\frac{7\pi}{4}}}\right\}$, respectively. The system $A$ is kept by Alice and $A'$ is sent to Bob via a quantum channel. Thus, the final state is given by
\begin{equation}
\rho_{AB}=(\mathbb{I}_A \otimes \mathcal{E}_{A'\rightarrow B})(\ket{\Phi}\bra{\Phi})_{AA'},
\end{equation}
where $\mathcal{E}_{A'\rightarrow B}$ is a completely positive and trace-preserving mapping. This mapping includes the influence of the environment and attacks by Eve. Alice randomly projects the system $A$ to an eigenstate by $\{M^x\}$. Then, she can write down the corresponding label $x$ without performing any additional operations. Bob measures the system $B$ as with the step \textit{(2)} in our prepare-and-measure protocol and follows the remaining steps.

\subsection{Key rate formula}
Applying reverse reconciliation means that Alice corrects her string $\textbf{X}$ according to Bob's string $\textbf{Z}$, which implies that adversary Eve should derive Bob's string unscrupulously. In the case where Eve performs collective attacks in the asymptotic regime, the secret key rate formula~\cite{lin2019asymptotic,devetak2005distillation} is given by
\begin{equation}
\begin{aligned}
R^{\infty} = p_{\rm pass}\min_{\rho_{AB}\in \textbf{S}}H(\textbf{Z}|E) - p_{\rm pass}\delta_{\rm EC},
\end{aligned}
\end{equation}
where $p_{\rm pass}$ is the sifting probability of preserving a round in the post-selection step to generate a raw key. The conditional von Neumann entropy $H(\textbf{Z}|E)$ describes the uncertainty of the string $\textbf{Z}$ from Eve's perspective. Eve's maximal knowledge of Bob's string $\textbf{Z}$ leads to the minimum uncertainty of $\textbf{Z}$ under certain density matrix $\rho_{AB}$. To evaluate Eve's maximal knowledge when $\rho_{AB}$ is uncertain for Alice and Bob, we must find the minimum conditional entropy $H(\textbf{Z}|E)$ of all $\rho_{AB}$ that conform to constraints $\textbf{S}$. $\delta_{\rm EC} $ is the amount of information leakage of each round in the error correction step. Considering the reconciliation efficiency, the leakage is
\begin{equation}
\begin{aligned}
\delta_{\rm EC}=&H(\textbf{Z})-\beta I(\textbf{X};\textbf{Z})\\
=&(1-\beta)H(\textbf{Z})+\beta H(\textbf{Z}|\textbf{X}),
\end{aligned}
\end{equation}
where $H(\textbf{Z})$ is the total information of $\textbf{Z}$ related to the probability distribution of Bob's measurement outcomes. $I(\textbf{X};\textbf{Z})$ is the classical mutual information between two strings. $H(\textbf{Z}|\textbf{X})$ is the conditional entropy describing the uncertainty of the string $\textbf{Z}$ conditioned on knowing the string $\textbf{X}$. The parameter $\beta$ represents the efficiency of error correction.

Because Bob announces the measurement quadrature of each round, we can analyze the security of each quadrature separately. To this end, we rewrite the secret key rate formula as
\begin{equation}
\begin{aligned}
R^{\infty} = \frac{1}{2}\sum_{y\in\{q,p\}}p^y_{\rm pass}\Big\{\min_{\rho_{AB}\in \textbf{S}}H(\textbf{Z}_y|E) - \delta^y_{\rm EC}\Big\}.
\end{aligned}\label{key0}
\end{equation}
The coefficient $\frac{1}{2}$ indicates that one half of states are measured in $\hat{q}$ and one half in $\hat{p}$. The secret key rate comprises the secret key rate in $\hat{q}$ and the secret key rate in $\hat{p}$. Note that Eve manipulates the density matrix $\rho_{AB}$ before Bob measures it, and she thus has no knowledge of Bob's choice of quadrature at that time. We can move the summation sign into the minimization problem because $\rho_{AB}$ of different quadratures are shared.

According to Refs.~\cite{coles2016numerical,winick2018reliable}, the conditional entropy terms in formula~(\ref{key0}) with their coefficients can be reformulated as
\begin{equation}
\frac{1}{2}\min_{\rho_{AB}\in \textbf{S}} \sum_{y\in\{q,p\}} D(\mathcal{G}_y(\rho_{AB})||\mathcal{Z}[\mathcal{G}_y(\rho_{AB})]),
\end{equation}
where $D(\rho||\sigma) = \text{Tr}(\rho \log_2\rho)-\text{Tr}(\rho \log_2\sigma)$ is the quantum relative entropy, $\mathcal{G}_{y}$ describes the post-processing of different quadratures, and $\mathcal{Z}$ is a pinching quantum channel that reads out key information. The post-processing mapping $\mathcal{G}_y(\rho) = K_y\rho K^{\dagger}_y$ corresponds to the measurement $\hat{y}\in\{\hat{q},\hat{p}\}$ and is given by
\begin{equation}\label{ky}
K_y = \sum^{1}_{b=0}\ket{b}_R\otimes \mathbb{I}_A\otimes(\sqrt{I_y^b})_B.
\end{equation}
Here $\ket{b}_R$ is the state of the key register $R$, which is determined by interval operators $\{I_y^b\}$. $I_y^b$ projects the system $B$ to one of two subspaces spanned by the eigenstates of operator $\hat{y}$ in terms of the mapping rule:
\begin{equation}
I_y^0 = \int^{\infty}_{\Delta}dy\ket{y}\bra{y} \text{, } I_y^1 = \int^{-\Delta}_{-\infty}dy\ket{y}\bra{y}.
\end{equation}
$\mathcal{Z}(\rho) = \sum^{1}_{b=0}Z_b\rho Z_b$ reads out the key from the mapping $\mathcal{G}_{y}(\rho_{AB})$ with $Z_0 = \ket{0}\bra{0}_R\otimes \mathbb{I}_{AB}$ and $Z_1 = \ket{1}\bra{1}_R\otimes \mathbb{I}_{AB}$. Here $\mathcal{G}_y$ is in a simplified form introduced by Ref.~\cite{lin2019asymptotic}. Kraus operators $\{K_y\}$ describe the part of the measurement and announcement: Alice's operation is described by an identity matrix because Alice announces nothing and has no effect on the register $R$. Her measurement in the entanglement-based protocol can be moved after the announcement and never influences $K_y$. Bob would announce the choice of quadrature, and his measurement is represented by interval operators. When considering the security, we can discuss two quadratures separately, denoted by the parameter $y\in\{q,p\}$ in the above formulas.

Finally, the key rate formula reads
\begin{equation}\label{key}
\begin{aligned}
R^{\infty} =& \frac{1}{2}\bigg\{\min_{\rho_{AB}\in \textbf{S}}\sum_{y\in\{q,p\}} D(\mathcal{G}_y(\rho_{AB})||\mathcal{Z}[\mathcal{G}_y(\rho_{AB})])\\
&-\sum_{y\in\{q,p\}}p^y_{\rm pass}\delta^y_{\rm EC}\bigg\}.
\end{aligned}
\end{equation}

\subsection{Numerical method}

\begin{figure}[t]
\centering
\includegraphics[width=8.6cm]{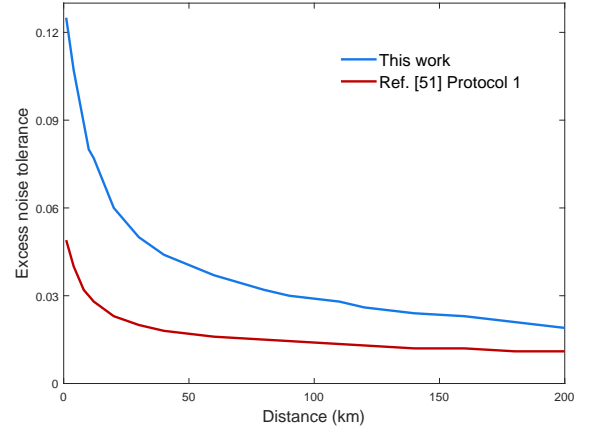}
\caption{Comparison between the excess noise tolerance of our protocol and that of Protocol 1 in Ref.~\cite{lin2019asymptotic}. Protocol 1 is a quaternary-modulated homodyne detection protocol that is similar to our protocol in terms of its implementation. The upper blue line is the performance of our protocol. The bottom red line is the performance of Protocol 1. The reconciliation efficiency is $\beta = 0.95$ and the post-selection is $\Delta = 0$. We plot this figure with the photon number cutoff $N_c=12$ and the maximal $N_i=100$ iterations of the first step in the numerical method.}\label{fig3}
\end{figure}

To realize the calculation of secret key rates, a photon-number cutoff assumption~\cite{lin2019asymptotic} is adopted. Based on this assumption, we can describe operators and density matrices in the photon-number representation with finite dimensions $N_c$. Because $\delta_{\rm EC}$ and $p_{\rm pass}$ are only concerned with the information held by Alice and Bob, the term $\sum_{y\in\{q,p\}}p^y_{\rm pass}\delta^y_{\rm EC}$ in the key rate formula~(\ref{key}) can be easily calculated according to general definitions~\cite{lin2019asymptotic}. We calculate the rest of the formula numerically. The rest is a minimization problem that can be described by
\begin{equation}\label{min}
\begin{aligned}
\text{minimize }&\sum_{y\in\{q,p\}} D(\mathcal{G}_y(\rho_{AB})||\mathcal{Z}[\mathcal{G}_y(\rho_{AB})])\\
\text{subject to}&\\
&\text{Tr}[\rho_{AB}(\ket{x}\bra{x}_A \otimes \hat{q})] = p_x\langle\hat{q}\rangle_x,\\
&\text{Tr}[\rho_{AB}(\ket{x}\bra{x}_A \otimes \hat{p})] = p_x\langle\hat{p}\rangle_x,\\
&\text{Tr}[\rho_{AB}(\ket{x}\bra{x}_A \otimes \hat{n})] = p_x\langle\hat{n}\rangle_x,\\
&\text{Tr}[\rho_{AB}(\ket{x}\bra{x}_A \otimes \hat{d})] = p_x\langle\hat{d}\rangle_x,\\
&\text{Tr}_B[\rho_{AB}] = \sum^{3}_{i,j=0}\sqrt{p_i p_j}\langle\phi_j|\phi_i\rangle\ket{i}\bra{j}_A,\\
&\text{Tr}[\rho_{AB}] = 1,\\
&\rho_{AB}\geqslant 0.
\end{aligned}
\end{equation}
The variable is the density matrix $\rho_{AB}$ subject to constraints $\textbf{S}$~\cite{lin2019asymptotic}. The first four constraints come from the experimental outcomes, where $x$ belongs to $\{00, 10, 11, 01\}$ and $\langle\hat{q}\rangle_x$, $\langle\hat{p}\rangle_x$, $\langle\hat{n}\rangle_x$, and $\langle\hat{d}\rangle_x$ are expectation values of operators when Bob measures states labeled by $x$. Homodyne detection directly outputs the outcomes of operators $\hat{q}$ and $\hat{p}$, while $\hat{n}=\frac{1}{2}(\hat{q}^2+\hat{p}^2-1)$ and $\hat{d} = \hat{q}^2-\hat{p}^2$ correspond to the second moments of $\hat{q}$ and $\hat{p}$. The next constraint about partial trace of system $B$ comes from the requirement of the completely positive and trace-preserving mapping, which implies that the quantum channel cannot influence the system $A$ of Alice. The last two constraints are natural requirements because $\rho_{AB}$ is a density matrix.

This minimization problem can be solved using many methods. We adopt one using a linearization method, given in Appendix~\ref{appA}. Briefly speaking, the numerical method involves two steps. First, we approach the optimal value of the minimization problem~(\ref{min}) by at most $N_i$ iterations. Second, the dual problem of the minimization problem is considered to guarantee that the result is less than or equal to the optimal value.

\section{PERFORMANCE}\label{sec:sim}
\begin{figure}[t]
\centering
\includegraphics[width=8.6cm]{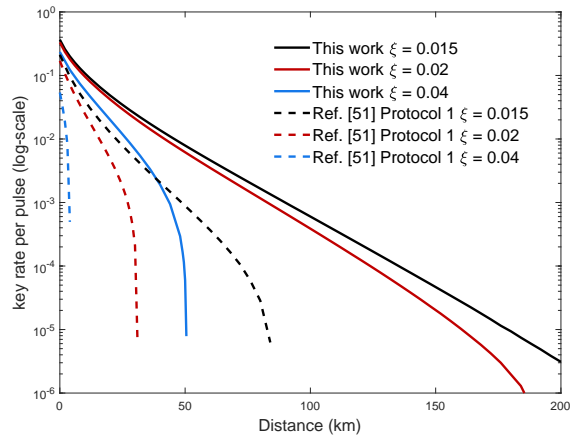}
\caption{Comparison between the secret key rate of our protocol and that of Protocol 1 in Ref.~\cite{lin2019asymptotic} with quaternary modulation and homodyne detection. From top to bottom, the solid lines represent the performance of our protocol under excess noise $\xi = 0.015,0.02,0.04$. The amplitude of our protocol is optimized in the range $[0.6,1.1]$ with step of $0.01$. From top to bottom, the dashed line represents the performance of Protocol 1 under excess noise $\xi = 0.015,0.02,0.04$. The signal state amplitude of Protocol 1 is optimized in range $[0.35, 0.6]$ with steps of $0.01$. For all points, $\beta = 0.95$ and $\Delta = 0$ are used. We plot this figure with $N_c=12$ and $N_i=300$.}\label{fig4}
\end{figure}

\begin{figure}[t]
\centering
\includegraphics[width=8.6cm]{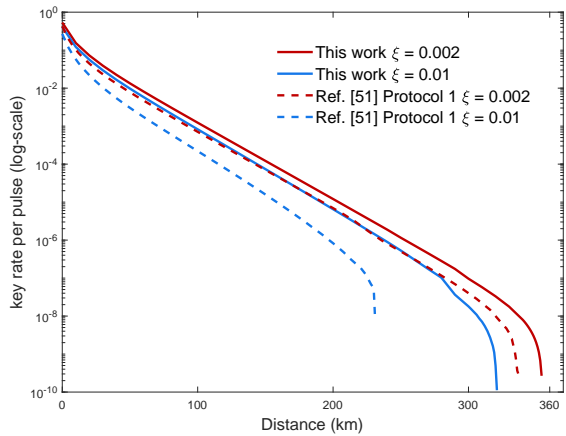}
\caption{Comparison between the secret key rate of our proposed protocol and that of Protocol 1 in Ref.~\cite{lin2019asymptotic} with quaternary modulation and homodyne detection. Solid lines from top to bottom show the performance of our protocol under excess noise $\xi=0.002, 0.01$. The amplitude of our protocol is optimized in the range $[0.6,1.1]$ with steps of $0.01$. Dashed lines from top to bottom are the performance of Protocol 1 under excess noise $\xi=0.002, 0.01$. The amplitude of Protocol 1 is optimized in the range $[0.35, 0.6]$ with steps of $0.01$. For all points, $\beta = 0.95$ and $\Delta = 0$ are used. We plot this figure with $N_c=12$ and $N_i=300$.}\label{fig5}
\end{figure}

Here, we investigate the performance of our protocol by performing numerical simulations. The simulation method is described in Appendix~\ref{appB}, where we set the reconciliation efficiency $\beta = 0.95$.

To simulate the performance, we adopt the phase-invariant Gaussian channel model in the absence of Eve~\cite{lin2019asymptotic}, which adds an effective excess noise to the input of the channel and transmits the input under the loss-only scenario. In this channel, the transmittance is $\eta = 10^{-\frac{aL}{10}}$ for distance $L$ with $a=0.2$ dB/km, and the effective excess noise is $\xi=\frac{\left(\mathrm{\Delta}q_{obs}\right)^2}{\left(\mathrm{\Delta}q_{vac}\right)^2}-1$. $\left(\mathrm{\Delta}q_{obs}\right)^2$ is the effective variance of $\hat{q}$ quadrature for the input of the channel, including the influence of detection noises. $\left(\mathrm{\Delta}q_{vac}\right)^2=1/2$ is the variance of the vacuum state in the $\hat{q}$ quadrature. Both variances are in the natural unit.
Because $\xi$ is normalized by $\left(\mathrm{\Delta}q_{vac}\right)^2$, the excess noise is under the shot-noise unit without vacuum noise.

We first emphasize the significant performance improvement of our proposed protocol by comparing it with Protocol 1 in Ref.~\cite{lin2019asymptotic}, which also prepares four types of coherent states and performs homodyne detection without post-selection.

As shown in Fig.~\ref{fig3}, we investigate the excess noise tolerance at different distances. The excess noise tolerance is calculated by maximizing the excess noise $\xi$ that enables the protocol to generate keys. The cutoff photon number $N_c$ is truncated at $12$ with $N_i=100$ iterations in the first step of the numerical method. The excess noise tolerance of our protocol is approximately double that of Protocol 1.

As shown Fig.~\ref{fig4}, we investigate the best secret key rates of our protocol for different levels of excess noise by optimizing the amplitude $\alpha$ of signal states in the interval $[0.6,1.1]$ with steps of $0.01$. We also calculate the best secret key rates of Protocol 1 by optimizing the amplitude $\alpha$ of signal states in the interval $[0.35, 0.6]$ with steps of $0.01$. Different search ranges are used because the optimal range of $\alpha$ varies for different protocols. The cutoff photon number $N_c$ is $12$ and the maximal iteration number $N_i$ of the first step in numerical method is $300$. The solid lines show the performance of our proposed protocol, whereas the dashed lines show the performance of Protocol 1. When the excess noise is experimentally feasible, such as $\xi = 0.02$, our protocol enables the distribution of keys for around 200 km with meaningful $10^{-6}$ bit secret keys per pulse. Moreover, when Protocol 1 can hardly generate keys for $\xi = 0.04$, our protocol can still distribute secret keys over 50 km.

\begin{figure}[t]
\centering
\includegraphics[width=8.6cm]{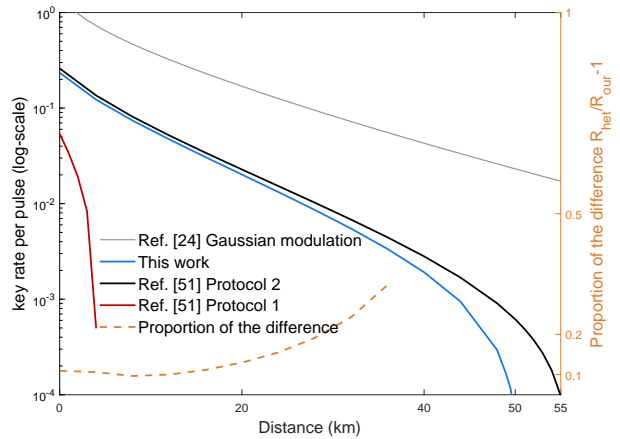}
\caption{Comparison of different CV-QKD protocols for the excess noise $\xi=0.04$. From top to bottom, solid lines represent the secret key rates of the Gaussian-modulated homodyne detection protocol~\cite{lodewyck2007quantum} with optimal modulated variance, the quaternary-modulated heterodyne detection protocol (Protocol 2)~\cite{lin2019asymptotic} with optimal amplitude of signals in the range $[0.6,1.1]$, this work with optimal amplitude in range $[0.6,1.1]$ and the quaternary-modulated homodyne detection protocol (Protocol 1)~\cite{lin2019asymptotic} with optimal amplitude in the range $[0.35, 0.6]$. The step of searching the optimal amplitude in the discrete-modulated protocol is 0.01, with $N_c=12$ and $N_i=300$. All CV-QKD protocols take the ideal detector scenario under consideration without post-selection and are against the collective attacks under the asymptotic regime. The difference between the secret key rates of this work and Protocol 2 is normalized by the secret key rate of this work $R_{het}/R_{our}-1$ and is shown with the y axis on the right.}
\label{fig6}
\end{figure}

\begin{figure}[t]
\centering
\includegraphics[width=8.6cm]{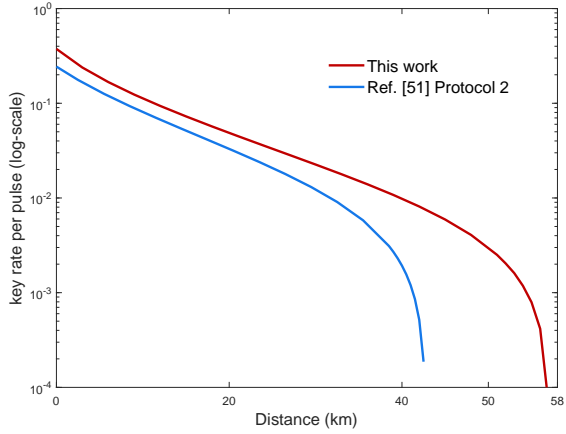}
\caption{Comparison of heterodyne detection protocol and this work under the untrusted detector noise scenario. The upper line is the secret key rate of this work, and the lower line is the secret key rate of Protocol 2 with heterodyne detection~\cite{lin2019asymptotic}. The excess noise of our protocol is $\xi=\xi_{oth}+\frac{\xi_{hom}}{\eta}$, and the excess noise of the heterodyne protocol is $\xi=\xi_{oth}+\frac{2\xi_{hom}}{\eta\times10^{-\frac{0.7dB}{10}}}$, where $\xi_{hom}=0.002$ is the detection noise of one homodyne detector, and $\xi_{oth}=0.01$ is the excess noise except detection noise. $\eta=10^{-\frac{0.2dB/km\times L}{10}}$ is the transmittance with distance $L$. We search the optimal amplitude for both protocols in the range $[0.6,1.1]$ with steps of $0.01$. For all points, $\beta = 0.95$ and $\Delta = 0$ are used. We plot this figure with $N_c=12$ and $N_i=100$.}
\label{fig7}
\end{figure}

In Fig.~\ref{fig5}, we show that there is an improvement in the transmission distance for a small excess noise with $N_c=12$ and $N_i=300$. When the excess noise is sufficiently small, but still possible, our protocol has the potential to distribute secret keys over 350 km. The transmission distance and secret key rates are increased significantly, which makes our protocol applicable and useful in the quantum secure communication network.

We further compare our protocol with a well-performed quaternary-modulated heterodyne detection protocol called Protocol 2 in Ref.~\cite{lin2019asymptotic}. Owing to differences in the detection methods, the comparison is made under two scenarios: ideal detector scenario and untrusted detector noise scenario.

Under the ideal detector scenario, both excess noise and loss are from the channel. We set the detection (electronic) noise of detectors $\xi_{det}=0$ and the detection efficiency $\eta_{det}=1$. In Fig.~\ref{fig6}, our protocol and Protocol 2 are simulated with parameters $\xi = 0.04$, $N_c=12$ and $N_i=300$. We search the best secret key rates of our protocol and Protocol 2 by optimizing the amplitude $\alpha$ of signal states in the interval $[0.6,1.1]$ with steps of $0.01$. Protocol 2 can generate a higher secret key rate compared with our proposed protocol under the ideal detector scenario because heterodyne detection can accumulate twice the amount of raw key data than homodyne detection. We call the secret key rate of Protocol 2 $R_{het}$ and the secret key rate of our protocol $R_{our}$. Then, we use $R_{het}/R_{our}-1$ to represent the proportion of the secret key rate difference $R_{het}-R_{our}$ in $R_{our}$, as shown by the dashed line in Fig.~\ref{fig6}. The secret key rate of Protocol 2 is no more than $20 \%$ higher than that of our proposed protocol within 30 km. The uncertainty relation limits the accuracy of measuring both quadratures simultaneously and increases the bit error rate of Protocol 2. Therefore, the improvement of the secret key rate by heterodyne detection is not significant. We also show the secret key rate of the Gaussian-modulated homodyne detection protocol in Fig.~\ref{fig6}. The modulated variance in the Gaussian-modulated protocol is the parameter that has been optimized.

In Fig.~\ref{fig7}, we regard the detector as a noisy and lossy device, and Eve can control the imperfection of the device, which is a more practical condition. We should note that the heterodyne detector often comprises two homodyne detectors and one 50:50 beam splitter that splits signal pulses. Therefore, two points of imperfection are considered under the untrusted detector noise scenario that we used here. First, the detection noise of a homodyne detector is $\xi_{hom}=0.002$, and the detection noise of a heterodyne detector is at least $\xi_{het}=2\xi_{hom}=0.004$. Second, the imperfection in the beam splitter usually causes an additional insertion loss of about $0.7$ dB. By setting other excess noise as $\xi_{oth}=0.01$, the total excess noise of our protocol is $\xi=\xi_{oth}+\frac{\xi_{hom}}{\eta}$, and the total excess noise of Protocol 2 with heterodyne detection is $\xi=\xi_{oth}+\frac{2\xi_{hom}}{\eta\times10^{-\frac{0.7}{10}}}$. The cutoff photon number $N_c$ is $12$ with $N_i=100$ iterations in the first step of the numerical method. Under this untrusted detector noise scenario, our protocol can perform better than the heterodyne detection protocol.

\begin{figure}[t]
\centering
\includegraphics[width=8.6cm]{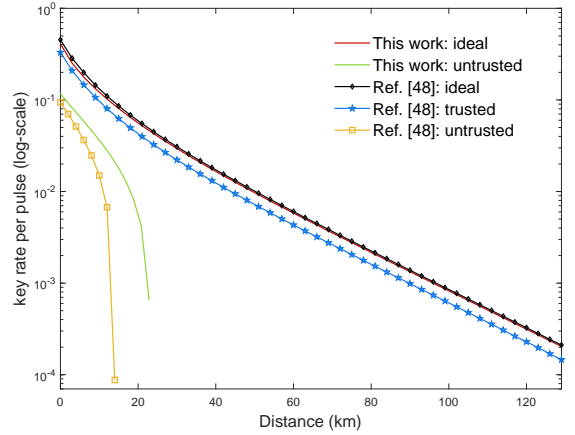}
\caption{Comparison of the heterodyne detection protocol~\cite{lin2019asymptotic,lin2020trusted} and this work under different detector noise scenarios. The upper two lines are under the ideal detector scenario, and the one for this work is a little lower. The lower two lines are under the untrusted detector noise scenario, and that for this study is higher. The performance of the heterodyne detection protocol under the trusted detector noise scenario~\cite{lin2020trusted} is the blue line in the middle. The data of the heterodyne protocol has been reported by the authors of Ref.~\cite{lin2020trusted}. The parameters are the same as those used in Fig. 3 in Ref.~\cite{lin2020trusted}. Specifically, the imperfect detector has excess noise $\xi_{hom}=0.01$ and detection efficiency $\eta_d = 0.719$. $\xi_{oth}=0.01$ is the excess noise and does not include detection noise. The total excess noise of our protocol under the untrusted detector noise scenario is $\xi=\xi_{oth}+\frac{\xi_{hom}}{\eta\times\eta_d}$. We search the optimal amplitude for our protocol in the range $[0.6,1.1]$ with steps of $0.01$. For all points, $\beta = 0.95$ and $\Delta = 0$ are used. We plot this figure with $N_c=15$ and the iteration of our protocol is $N_i=100$.}
\label{fig8}
\end{figure}

In Fig.~\ref{fig8}, we compare the heterodyne protocol~\cite{lin2019asymptotic,lin2020trusted} under ideal, trusted, and untrusted detector noise scenarios with our protocol under ideal and untrusted detector noise scenarios. The data of the heterodyne protocol were reported by the authors of Ref.~\cite{lin2020trusted}. The imperfect detector has excess noise $\xi_{hom}=0.01$ and detection efficiency $\eta_d = 0.719$. The other excess noise is $\xi_{oth}=0.01$ and the cutoff photon number $N_c$ is $15$. We simulate our proposed protocol using the same parameters. Thus, the total excess noise of our protocol under the untrusted detector noise scenario is $\xi=\xi_{oth}+\frac{\xi_{hom}}{\eta\times\eta_d}$. The iteration in the first step of the numerical method is $N_i=100$. Our protocol is a little worse than the heterodyne protocol under the ideal scenario, but it is better under the untrusted detector noise scenario. The heterodyne protocol with trusted detector noise has a key rate that is comparable to the same protocol with the ideal detector, which may imply good performance when we also consider our protocol under the trusted detector noise scenario.

\begin{figure}[t]
\centering
\includegraphics[width=8.6cm]{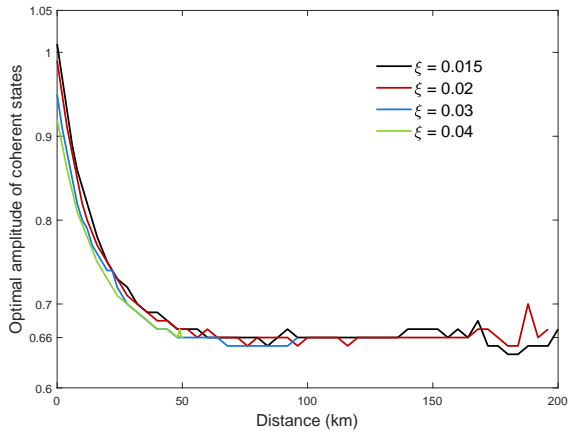}
\caption{Optimal amplitudes of the coherent states prepared by Alice for different transmission distances.
From top to bottom, the lines correspond to the conditions where the excess noises are $\xi = 0.015, 0.02, 0.03, 0.04$. For all points, $\beta = 0.95$ and $\Delta = 0$ are used. We plot this figure with $N_c=12$ and $N_i=300$.}
\label{fig9}
\end{figure}

For the experiments, we show the optimal amplitudes that are searched for different transmission distances with parameters $N_c=12$ and $N_i=300$. As shown in Fig.~\ref{fig9}, in the case of a long distance, the optimal amplitude decreases as the transmission distance increases, except for the case when there are some jitters, and a larger excess noise induces a lower optimal amplitude with little violation. The optimal amplitude for remote users is about 0.66. This conclusion also applies to the condition where $N_c=12$ and $N_i=100$.

\begin{figure}[t]
\centering
\includegraphics[width=8.6cm]{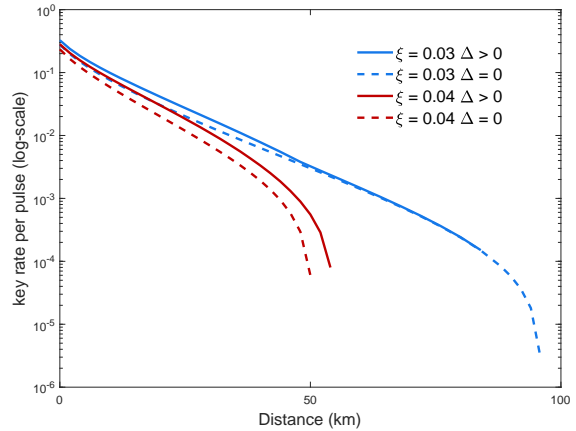}
\caption{Performance of the optimal post-selection. Cases when $\xi=0.03$ and $\xi=0.04$ are considered. The top solid line in blue represents the performance of the post-selection ($\Delta > 0$) with $\xi=0.03$, while the top dashed line in blue is the performance without post-selection. The lower solid line in red represents the performance of the post-selection $\xi=0.04$, while the lower dashed line in red is the performance without post-selection ($\Delta=0$). The amplitude for the point in the dashed lines is optimized in the range $[0.6,1.1]$ with steps of $0.01$. The amplitude for the point in the solid lines is the same with dashed lines under the same distance and excess noise. For $\xi=0.04$, at the distance where our protocol is unable to generate keys, we use the amplitude $\alpha=0.66$ in post-selection. For all points, $\beta = 0.95$ is used. We plot this figure with $N_c=12$ and $N_i=100$.}
\label{fig10}
\end{figure}

The post-selection is a useful method to further increase the transmission distance and improve the secret key rate. In Fig.~\ref{fig10}, we optimize the sifting parameter $\Delta$ using the amplitude of signal states that have been optimized without post-selection. The cutoff photon number $N_c$ is truncated at $12$ with $N_i=100$ iterations. The available $\Delta$ has steps of 0.01. At distances where secret key rates are zero without post-selection, we search the best post-selection parameter $\Delta$ by setting the amplitude to 0.66. This choice is reasonable considering that the optimal amplitude for the long distance is around 0.66 in average, as depicted in Fig.~\ref{fig9}. The post-selection is more important for the scenario with the higher excess noise. For $\xi=0.04$, the secret key rates are improved, and the largest transmission distance is extended. For smaller excess noise $\xi=0.03$, the improvement of post-selection is subtle. Over a long distance, there is no improvement in the secret key rates for $\xi=0.03$. This is because as the excess noise increases, the overlap of outcome distributions obtained by measuring different kinds of signal states also increases, which causes higher bit error rates. The post-selection discards outcomes around zero because the overlap of outcome distributions most likely happens around zero. By post-selection, we discard many error bits, thus reducing the bit error rates. Protocols with heterodyne detection enjoy the same advantages of post-selection~\cite{kanitschar2021postselection}. The optimal $\Delta$ is larger under longer transmission distance. The sifting probability related to post-selection decreases to around 0.15 over the longest transmission distance for $\xi=0.04$; the post-selection thus makes sense in our protocol. This operation can work well when the amplitude cannot be prepared optimally in the experiment.

Note that the curves in the above figures should have been smooth in theory. Although the secret key rate formula is a convex function with convex domain $\rho_{AB}$, it is difficult to unify the imprecision of $\rho_{AB}$ under different amplitudes and transmittances. The inaccuracy also accumulates along with steps of calculations, resulting in some calculation noise. Finally, as a numerical method, limited by the calculation accuracy of computers and the stability of the convex optimization algorithm, unstable points and unsmooth curves appear. The quality of the algorithm largely determines the secret key rates. When the ideal secret key rate is quite small that it reaches the computational accuracy limit, numerical results sometimes show that no keys can be generated. Achieving improvements in the algorithm may help further enhance the transmission distance and stabilize the key rates. The calculation speed of the numerical method is often slow and is approximately 20 min per set of parameters. The speed largely depends on the number of iterations and the truncated photon number. Intriguingly, recent work~\cite{zhou2021machine} can utilize a neural network model to quickly predict the secure key rate of homodyne detection discrete-modulated CV-QKD with high accuracy.

\section{DISCUSSION}\label{sec:dis}
Homodyne detection is the first adopted measurement method of CV-QKD with noticeable performance experimentally, and it provides excellent tolerance in terms of excess noise in Gaussian-modulated protocols.
Our main goal is to inherit this mature technology and apply it to the discrete-modulated protocol. We propose a quaternary-modulated homodyne detection protocol with very high excess noise tolerance and long transmission distance that has never been reached by previous discrete-modulated homodyne detection protocols~\cite{PhysRevA.79.012307,PhysRevA.97.022310,lin2019asymptotic}. There is a quaternary-modulated heterodyne detection protocol~\cite{lin2019asymptotic} that shows high secret key rate and long transmission distance under large excess noise based on simulations. However, the heterodyne detector~\cite{PhysRevLett.93.170504,lance2005no,brunner2017low,chin2021machine,Jain_2021} is more sophisticated than the homodyne detector, and it introduces more noise and loss in experiments~\cite{RevModPhys.66.481,leonhardt1997measuring, PhysRevA.82.042315,CVreview}. We show that our proposed protocol can offer a comparable performance with this heterodyne detection protocol under the ideal detector scenario. Our protocol can even perform better than this heterodyne detection protocol when considering the untrusted detector noise condition.

We determine why the previous quaternary-modulated homodyne detection protocol~\cite{lin2019asymptotic} cannot transmit over a long distance although they also send four kinds of states. In particular, from a practical point of view, our protocol and the previous protocol both generate one bit per pulse by homodyne detection. Compared with Eq. (17) in Ref.~\cite{lin2019asymptotic}, the mapping $K_y$ of our protocol in Eq.~(\ref{ky}) is an identity matrix at Alice's end, which means that Alice remains silent in our protocol. This mathematical difference implies that additional announcements cause more information leakage.

In addition, the adoption of QPSK is required during the preparation.
Note that QPSK is a mature method in classical communication, shifting the phase of states by $\{\frac{\pi}{4}, \frac{3\pi}{4}, \frac{5\pi}{4}, \frac{7\pi}{4}\}$.
In previous quaternary-modulated protocols~\cite{ghorai2019asymptotic,lin2019asymptotic}, they use states with phases $\{0, \frac{\pi}{2}, \pi, \frac{3\pi}{2}\}$.
The probability distribution of a state in quadrature $\hat{p}$ is symmetric about the original point if the state has phase $0$ or $\pi$. Thus, states with phase $0$ or $\pi$ cannot represent keys when Bob applies $\hat{p}$ measurement. Similarly, states with $\frac{\pi}{2}$ or $\frac{3\pi}{2}$ cannot represent keys when Bob applies $\hat{q}$. The additional $\frac{\pi}{4}$ phase shift in our protocol offers chances to generate keys in both quadratures, and Alice can thus announce nothing.

Eventually, owing to the tolerance of excess noise, the transmission distance is improved, which removes the obstruction of the proceedings of the large-scale discrete-modulated CV-QKD network. It is worth recalling that commercial instruments for QPSK have been updated to over 50 Gbps~\cite{po201250gbs}, which implies that the speed of preparation in our proposed protocol can also be increased to 50 Gbps. Our protocol will maximize the use of the mature homodyne detection technology to achieve the long-distance key distribution. The experimental demonstration of our protocol is of great practical value for network security and information security. Moreover, we can consider the trusted detector noise model~\cite{lin2020trusted}. With a simpler preparation, our proposed protocol has a chance to exceed the transmission record of 202.81 km set by the Gaussian modulation protocol~\cite{zhang2020long} with an ultralow-loss fiber and small excess noise parameters. The experimental schematic and the search of optimal amplitudes in the text provide instructions for the experimental implementation.

Further studies of our protocol's security are also desirable. Note that the photon-number cutoff assumption can be removed~\cite{upadhyaya2021dimension} in the heterodyne detection protocol, which inspires us to present a more rigorous security analysis of our protocol in future work. Moreover, we can focus on determining a suitable adjustment of security theory to deal with the finite-size effects. The security against coherent attacks is another interesting and important direction for discrete-modulated CV-QKD. The quantum de Finetti representation theorem~\cite{renner2008security} is a possible path to equalize the information leaked by coherent attacks and collective attacks.

\section*{Acknowledgments}

We thank Jie Lin for valuable discussions on the numerical method in Ref.~\cite{lin2019asymptotic} and providing data points from Ref.~\cite{lin2020trusted} for the comparison in Fig.~\ref{fig8}. We gratefully acknowledge the support received from the National Natural Science Foundation of China (under Grant No. 61801420), the Key-Area Research and Development Program of Guangdong Province (under Grant No. 2020B0303040001), and the Fundamental Research Funds for the Central Universities (under Grant No. 020414380182).

\appendix
\section{Numerical methods of calculating the key rates}\label{appA}
The minimization problem in Eq.~(\ref{min}) is a convex optimal problem with one unique solution. It can be described as follows.

The independent variable is the density matrix $\rho_{AB}$.
The dimension number of Alice's system is $4$, which is determined by the number of different states she prepared. $\ket{x}\bra{x}_A$ with $x\in \{00, 10, 11, 01\}$ projects Alice's system into a one-dimensional subspace, which implies that Alice sends state $\ket{\phi_x}\in\left\{\ket{\alpha e^{i\frac{\pi}{4}}}, \ket{\alpha e^{i\frac{3\pi}{4}}}, \ket{\alpha e^{i\frac{5\pi}{4}}}, \ket{\alpha e^{i\frac{7\pi}{4}}}\right\}$ to Bob. However, the state that Bob receives is infinite-dimensional. The photon-number cutoff assumption is proposed to cut off the dimension of Bob's system. According to the assumption, $\rho_{AB}$ can be represented by a finite-dimensional matrix in Fock representation. Thus, the numerical method can be applied. The operators $\hat{q}, \hat{p}, \hat{n}, \hat{d}$ correspond to the measurement on Bob's system. Considering the photon-number representation, they can be defined by the annihilation operator $\hat{a}$:
\begin{equation}
	\begin{aligned}
		\hat{q}&=\frac{1}{\sqrt{2}}(\hat{a} + \hat{a}^\dagger),\\
		\hat{p}&=\frac{i}{\sqrt{2}}(\hat{a}^\dagger - \hat{a}),\\
		\hat{n}&=\hat{a}^\dagger \hat{a} = \hat{a}\hat{a}^\dagger - 1,\\
		\hat{d}&= \hat{a}^2 + (\hat{a}^\dagger)^2.
	\end{aligned}
\end{equation}
To calculate them numerically, the matrix of $\hat{a}$ in photon-number representation is also cut off according to the photon-number cutoff assumption. We introduce notation $\langle\hat{q}\rangle_x, \langle\hat{p}\rangle_x, \langle\hat{n}\rangle_x, \langle\hat{d}\rangle_x$ to represent the expectation values of these operators.
The relative entropy $D(\rho||\sigma)$ is the dependent variable that we need to minimize. $\mathcal{G}(\rho)=K\rho K^{\dagger}$ is the map in post-processing with Kraus operator $K$.

To implement the convex optimization, a two-step numerical method is introduced~\cite{winick2018reliable} below. For convenience, we use $f(\rho)$ to represent the function $D(\mathcal{G}(\rho)||\mathcal{Z}[\mathcal{G}(\rho)])$. The constraints can be described by a set $\textbf{S}$:
\begin{equation}
	\textbf{S} = \{\rho \in \mathcal{H}_{+} | \text{Tr}(\Gamma_i \rho) = \gamma_i , \forall i\},
\end{equation}
where $\mathcal{H}_{+}$ is the set of positive semidefinite operators. $\Gamma_i$ is the Hermitian operator and $\gamma_i$ is the corresponding expectation value. To fit this form, one should rewrite the fifth constraint in Eq.~(\ref{min}). Specifically, one needs to transform the partial trace on Bob into some Hermitian operators.

Using the linearization method, we can obtain
\begin{equation}\label{linear}
\begin{aligned}
f(\rho^*)\geqslant& f_{\epsilon}(\rho) + \min_{\rho + \Delta\rho \in \textbf{S}}[\text{Tr}((\Delta \rho)^T \nabla f_{\epsilon}(\rho))]-\zeta_{\epsilon},\\
=& f_{\epsilon}(\rho) - \text{Tr}(\rho^T \nabla f_{\epsilon}(\rho))+ \min_{\sigma\in \textbf{S}}[\text{Tr}(\sigma^T \nabla f_{\epsilon}(\rho))]-\zeta_{\epsilon},
\end{aligned}
\end{equation}
where $\rho^*$ is the optimal density matrix (i.e., the solution of minimization problem Eq.~(\ref{min}) and $\epsilon$ represents a perturbation parameter, by which the gradient of $f(\rho)$ can exist. Owing to the existence of $\epsilon$, the term $\zeta_{\epsilon} = 2\epsilon(d'-1)\log_2\frac{d'}{\epsilon(d'-1)}$ is applied to correct the difference between $f(\rho)$ and $f_{\epsilon}(\rho)$ caused by perturbation. In our simulation, we set $\epsilon = 10^{-12}$ to decrease the value of $\zeta_{\epsilon}$. Using $\epsilon$, the dependent variable becomes
\begin{equation}
f_{\epsilon}(\rho) = D(\mathcal{G}_{\epsilon}(\rho)||\mathcal{Z}[\mathcal{G}_{\epsilon}(\rho)]).
\end{equation}
The gradient of $f_{\epsilon}(\rho)$ is
\begin{equation}
[\nabla f_{\epsilon}(\rho)]^T = \mathcal{G}_{\epsilon}^{\dagger}(\log_2\mathcal{G}_{\epsilon}(\rho)) -\mathcal{G}_{\epsilon}^{\dagger}(\log_2\mathcal{Z}(\mathcal{G}_{\epsilon}(\rho))),
\end{equation}
where
\begin{equation}
\begin{aligned}
&\mathcal{G}_{\epsilon}(\rho) = (1-\epsilon)\mathcal{G}(\rho) + \epsilon \mathbb{I}/d',\\
&\mathcal{Z}(\mathcal{G}_{\epsilon}(\rho)) = (1-\epsilon)\mathcal{Z}(\mathcal{G}(\rho)) + \epsilon \mathbb{I}/d'.
\end{aligned}
\end{equation}
$d'$ is the dimension of the matrix $\mathcal{G}(\rho)$, which is twice the dimension of density matrix $\rho$.

Using the linearization expressed in Eq.~(\ref{linear}), the optimization problem is transformed into a semi-definite program, such as the second or third term of the first or the second line, respectively, in Eq.~(\ref{linear}). Therefore, we only need to find $\Delta \rho$ or $\sigma = \rho +\Delta \rho$ that minimizes the related term. We take $\Delta \rho$ as an example, and the first step is to find a density matrix $\rho$ that is quite close to the optimal one. To accomplish this step, we perform the algorithm as follows:\\
1. Begin with any $\rho_0 \in \textbf{S}$ and set i = 0.\\
2. Compute $\min_{\rho + \Delta\rho \in \textbf{S}}[\text{Tr}((\Delta \rho)^T \nabla f_{\epsilon}(\rho_i))]$.\\
3. If $\text{Tr}((\Delta \rho)^T \nabla f_{\epsilon}(\rho_i))$ is sufficiently small, STOP.\\
4. Find $\lambda\in(0,1)$ that minimizes $f_{\epsilon}(\rho_i + \lambda\Delta\rho)$.\\
5. Set $\rho_{i+1} = \rho_i + \lambda\Delta\rho$, $i \leftarrow i+1$ and go to 2.\\

When solving the minimization problem in step 2 of the algorithm, we find that the results are often sensitive to the setting of $\rho_0$. Different convergence paths affect the convergence effect. In our simulation, we often choose the result of feasible verification under $\textbf{S}$ or the result that maximizes the minimal eigenvalue of $\rho$ under $\textbf{S}$ as the primal density matrix $\rho_0$. The iteration of 2--5 sometimes needs to be broken manually; otherwise, it may take a large amount of time to converge or even never stop. The largest iteration number in our simulation is 300 or 100. After 300 or 100 iterations, we choose $\rho_i$ with the smallest $\text{Tr}((\Delta \rho)^T \nabla f_{\epsilon}(\rho_i))$ among all iterations as our result of the first step.

Reference~\cite{winick2018reliable} does not offer any method to deal with the minimization of $f_{\epsilon}(\rho_i + \lambda\Delta\rho)$ because its DV-QKD example is simple for calculation. However, it is not an easy task for CV-QKD. We provide our own method inspired by BinarySearch:\\
1. Set a starting point $\lambda_s = 0$, an ending point $\lambda_e = 1$, and a middle point $\lambda_m = 0.5$.\\
2. Calculate $f_{\epsilon}(\rho_i + \lambda\Delta\rho)$ in three points, and compare them.\\
3. If $\lambda_s$ is the smallest one, update the ending point by $\lambda_e = \lambda_m$ and the middle point by $\lambda_m = 0.5(\lambda_s+\lambda_e)$. Then, go to 2.\\
4. If $\lambda_e$ is the smallest one, update the starting point by $\lambda_s = \lambda_m$ and the middle point by $\lambda_m = 0.5(\lambda_s+\lambda_e)$. Then, go to 2.\\
5. If $\lambda_m$ is the smallest one, two temporary points are generated by $\lambda_1 = 0.5(\lambda_s+\lambda_m)$ and $\lambda_2 = 0.5(\lambda_m+\lambda_e)$.\\
6. Calculate $f_{\epsilon}(\rho_i + \lambda\Delta\rho)$ of two temporary points and compare them with $\lambda_m$.\\
7. If $\lambda_m$ is still the smallest one, set $\lambda_s = \lambda_1$ and $\lambda_e = \lambda_2$, and go to 2.\\
8. If $\lambda_1$ is the smallest one, set $\lambda_e = \lambda_m$ and $\lambda_m = \lambda_1$, and go to 2.\\
9. If $\lambda_2$ is the smallest one, set $\lambda_s = \lambda_m$ and $\lambda_m = \lambda_2$, and go to 2.\\

The calculation results of some points can be temporarily stored for the next iteration. The termination of the iteration is decided by the calculation accuracy of computers.

After the first step, we seemingly find an appropriate density matrix $\rho$. However, considering the imprecision of numerical calculations, we cannot guarantee that the solution of the first step is the optimal one. In fact, we find an upper bound of the solution. To protect the security of the generated keys, we use the dual problem in the second step.

One can transform the primal problem $\min_{\sigma\in \textbf{S}}[\text{Tr}(\sigma^T \nabla f_{\epsilon}(\rho))]$ into the dual problem $\max_{\vec{y}\in\textbf{S}^*(\rho)}\vec{\gamma}\cdot \vec{y}$. The variable $\vec{y}$ obeys the constraints:
\begin{equation}
	\textbf{S}^*(\rho) = \{\vec{y} \in \mathbb{R}^{n} | \sum_i y_i \Gamma^T_i \leqslant \nabla f_{\epsilon}(\rho)\}.
\end{equation}
In general, primal and dual problems satisfy the weak duality. The weak duality shows that the optimal value of the dual problem is always less than or equal to the optimal value of the primal problem. According to this principle, the final solution that we obtain from the dual problem must be a lower bound of the primal problem even if it is not optimal. The lower bound of the primal problem is also the lower bound of the secret key rates according to Eq.~(\ref{linear}).

In addition, we can consider the imprecision of the floating-point representations of $\{\Gamma_i\}$ and $\{\gamma_i\}$.
\begin{equation}
	|\text{Tr}(\tilde{\Gamma}_i \rho) - \tilde{\gamma}_i| \leq \epsilon_{\rm rep},
\end{equation}
where $\{\tilde{\Gamma}_i\}$ and $\{\tilde{\gamma}_i\}$ are the approximate representations used in the calculation. Furthermore, the solution $\rho$ may not satisfy the constraints, especially the positive semi-definite constraint. Ref.~\cite{winick2018reliable} encourages one to transform $\rho$ into positive semi-definite matrix $\rho'$ by subtracting a term $\lambda_{\rm min}\mathbb{I}$ when the smallest eigenvalue $\lambda_{\rm min}$ of $\rho$ is negative. Then, we restrict the imprecision by
\begin{equation}
	|\text{Tr}(\tilde{\Gamma}_i \rho') - \tilde{\gamma}_i| \leq \epsilon_{\rm sol}.
\end{equation}
After the first step, we extract the quantity $\epsilon' = \max(\epsilon_{\rm rep},\epsilon_{\rm sol})$ from the output $\rho$. This quantity influences the second step. The dual problem solved in the second step is adjusted to
\begin{equation}
	\max_{(\vec{y},\vec{z})\in\tilde{\textbf{S}}^*(\rho)}\left(\vec{\gamma} \cdot \vec{y}-\epsilon ' \sum_{i=1}^{n}z_i\right),
\end{equation}
where
\begin{equation}
	\begin{aligned}
		&\tilde{\textbf{S}}^*(\rho) \\
		&:= \{(\vec{y},\vec{z}) \in (\mathbb{R}^{n},\mathbb{R}^{n}) | -\vec{z} \leqslant \vec{y} \leqslant \vec{z}, \sum_i^n y_i \tilde{\Gamma}
		^T_i \leqslant \nabla f_{\epsilon}(\rho)\}.\\
	\end{aligned}
\end{equation}
The description above is about one quadrature. Because $f_q(\rho)$ for $\hat{q}$ and $f_p(\rho)$ for $\hat{p}$ are added linearly, we can replace $f_{\epsilon}(\rho)$ by $f_{q,\epsilon}(\rho)+f_{p,\epsilon}(\rho)$. Then, $\nabla f_{\epsilon}(\rho)$ in primal and dual problems is replaced by $\nabla f_{q,\epsilon}(\rho)+\nabla f_{p,\epsilon}(\rho)$. In addition, the perturbation compensation $\zeta_{\epsilon}$ is multiplied by 2.

Thus far, the lower bound of the minimization problem is solved numerically. Note that we solve primal and dual problems using the SDPT3 solver in CVX 1.22 package on MATLAB R2020b.
This package is for specifying and solving convex programs, and SDPT3 is a free solver~\cite{cvx,gb08}. The code of our protocol is uploaded to the open-source code website~\cite{wherecode}.

\section{Simulation Methods}\label{appB}
We simulate the experiment's statistical results by assuming a phase-invariant Gaussian channel, as referred in Ref.~\cite{lin2019asymptotic}. When Alice sends a coherent state $\ket{\alpha}$, Bob obtains a displaced thermal state centered at $\sqrt{\eta}\alpha$ with the variance $\frac{1}{2}+\frac{\eta\xi}{2}$ for each quadrature. The first term $\frac{1}{2}$ in the variance is the vacuum noise. Thus, $\rho_B^x$ can be given by
\begin{equation}
\rho_B^x=\frac{1}{\pi V_A}\int e^{-|\beta-\sqrt{\eta}\alpha_x|^2/V_A}\ket{\beta}\bra{\beta} d^2\beta,
\end{equation}
with $\alpha_x \in \left\{\ket{\alpha e^{i\frac{\pi}{4}}}, \ket{\alpha e^{i\frac{3\pi}{4}}}, \ket{\alpha e^{i\frac{5\pi}{4}}}, \ket{\alpha e^{i\frac{7\pi}{4}}}\right\}$ and $V_A = \frac{\eta\xi}{2}$. $\xi$ is the excess noise and $\eta = 10^{-\frac{aL}{10}}$ is the transmittance with $a=0.2$ dB/km in the distance $L$. Thus, the expectation values of measurement operators are
\begin{equation}
\begin{aligned}
\langle \hat{q} \rangle_x&= \sqrt{2\eta}Re(\alpha_x),\\
\langle \hat{p} \rangle_x&= \sqrt{2\eta}Im(\alpha_x),\\
\langle \hat{n} \rangle_x&= \eta|\alpha_x|^2 + \frac{\eta\xi}{2},\\
\langle \hat{d} \rangle_x&= \eta[(\alpha_x)^2 +(\alpha^*_x)^2].
\end{aligned}
\end{equation}
To simulate the error correction, the conditional probability between Alice's sending state and Bob's mapping result can be given by
\begin{equation}
P(q|x) = \frac{1}{\sqrt{\pi(\eta\xi+1)}}e^{[-(q-\sqrt{2\eta}Re(\alpha_x))^2]/(\eta\xi+1)},
\end{equation}
to measure $\hat{q}$ and
\begin{equation}
P(p|x) = \frac{1}{\sqrt{\pi(\eta\xi+1)}}e^{[-(q-\sqrt{2\eta}Im(\alpha_x))^2]/(\eta\xi+1)},
\end{equation}
to measure $\hat{p}$.
According to the bit mapping of Bob, the probability distribution of the bit is
\begin{equation}
\begin{aligned}
P_y(0|x) &= \int^{\infty}_{\Delta} P(y|x)dy,\\
P_y(1|x) &= \int^{\Delta}_{-\infty} P(y|x)dy,\\
P_y(\bot|x) &= \int^{\Delta}_{-\Delta} P(y|x)dy,
\end{aligned}
\end{equation}
where $y\in\{q,p\}$.
Then, we can find the sifting probability
\begin{equation}
p^y_{\rm pass}= \sum_{x}p_x(P_y(0|x)+P_y(1|x)),
\end{equation}
and normalize the probability distribution after discarding bits $\bot$.
The conditional entropy for each quadrature is thus calculated by
\begin{equation}
H(\textbf{Z}_y|\textbf{X}_y) = \sum_x p_x h\left(\frac{P_y(0|x)}{P_y(0|x)+P_y(1|x)}\right),
\end{equation}
where $h(x)= -x\log_{2}{x}-(1-x)\log_{2}{(1-x)}$. Thus, the experimental results required for calculating the key rates can be simulated by utilizing the above equations.

%


\begin{thebibliography}{77}%
\makeatletter
\providecommand \@ifxundefined [1]{%
 \@ifx{#1\undefined}
}%
\providecommand \@ifnum [1]{%
 \ifnum #1\expandafter \@firstoftwo
 \else \expandafter \@secondoftwo
 \fi
}%
\providecommand \@ifx [1]{%
 \ifx #1\expandafter \@firstoftwo
 \else \expandafter \@secondoftwo
 \fi
}%
\providecommand \natexlab [1]{#1}%
\providecommand \enquote  [1]{``#1''}%
\providecommand \bibnamefont  [1]{#1}%
\providecommand \bibfnamefont [1]{#1}%
\providecommand \citenamefont [1]{#1}%
\providecommand \href@noop [0]{\@secondoftwo}%
\providecommand \href [0]{\begingroup \@sanitize@url \@href}%
\providecommand \@href[1]{\@@startlink{#1}\@@href}%
\providecommand \@@href[1]{\endgroup#1\@@endlink}%
\providecommand \@sanitize@url [0]{\catcode `\\12\catcode `\$12\catcode
  `\&12\catcode `\#12\catcode `\^12\catcode `\_12\catcode `\%12\relax}%
\providecommand \@@startlink[1]{}%
\providecommand \@@endlink[0]{}%
\providecommand \url  [0]{\begingroup\@sanitize@url \@url }%
\providecommand \@url [1]{\endgroup\@href {#1}{\urlprefix }}%
\providecommand \urlprefix  [0]{URL }%
\providecommand \Eprint [0]{\href }%
\providecommand \doibase [0]{https://doi.org/}%
\providecommand \selectlanguage [0]{\@gobble}%
\providecommand \bibinfo  [0]{\@secondoftwo}%
\providecommand \bibfield  [0]{\@secondoftwo}%
\providecommand \translation [1]{[#1]}%
\providecommand \BibitemOpen [0]{}%
\providecommand \bibitemStop [0]{}%
\providecommand \bibitemNoStop [0]{.\EOS\space}%
\providecommand \EOS [0]{\spacefactor3000\relax}%
\providecommand \BibitemShut  [1]{\csname bibitem#1\endcsname}%
\let\auto@bib@innerbib\@empty
\bibitem [{\citenamefont {Arute}\ \emph {et~al.}(2019)\citenamefont {Arute},
  \citenamefont {Arya}, \citenamefont {Babbush}, \citenamefont {Bacon},
  \citenamefont {Bardin}, \citenamefont {Barends}, \citenamefont {Biswas},
  \citenamefont {Boixo}, \citenamefont {Brandao}, \citenamefont {Buell} \emph
  {et~al.}}]{arute2019quantum}%
  \BibitemOpen
  \bibfield  {author} {\bibinfo {author} {\bibfnamefont {F.}~\bibnamefont
  {Arute}}, \bibinfo {author} {\bibfnamefont {K.}~\bibnamefont {Arya}},
  \bibinfo {author} {\bibfnamefont {R.}~\bibnamefont {Babbush}}, \bibinfo
  {author} {\bibfnamefont {D.}~\bibnamefont {Bacon}}, \bibinfo {author}
  {\bibfnamefont {J.~C.}\ \bibnamefont {Bardin}}, \bibinfo {author}
  {\bibfnamefont {R.}~\bibnamefont {Barends}}, \bibinfo {author} {\bibfnamefont
  {R.}~\bibnamefont {Biswas}}, \bibinfo {author} {\bibfnamefont
  {S.}~\bibnamefont {Boixo}}, \bibinfo {author} {\bibfnamefont {F.~G.}\
  \bibnamefont {Brandao}}, \bibinfo {author} {\bibfnamefont {D.~A.}\
  \bibnamefont {Buell}}, \emph {et~al.},\ }\bibfield  {title} {\bibinfo {title}
  {Quantum supremacy using a programmable superconducting processor},\
  }\href@noop {} {\bibfield  {journal} {\bibinfo  {journal} {Nature}\ }\textbf
  {\bibinfo {volume} {574}},\ \bibinfo {pages} {505} (\bibinfo {year}
  {2019})}\BibitemShut {NoStop}%
\bibitem [{\citenamefont {Zhong}\ \emph {et~al.}(2020)\citenamefont {Zhong},
  \citenamefont {Wang}, \citenamefont {Deng}, \citenamefont {Chen},
  \citenamefont {Peng}, \citenamefont {Luo}, \citenamefont {Qin}, \citenamefont
  {Wu}, \citenamefont {Ding}, \citenamefont {Hu} \emph
  {et~al.}}]{zhong2020quantum}%
  \BibitemOpen
  \bibfield  {author} {\bibinfo {author} {\bibfnamefont {H.-S.}\ \bibnamefont
  {Zhong}}, \bibinfo {author} {\bibfnamefont {H.}~\bibnamefont {Wang}},
  \bibinfo {author} {\bibfnamefont {Y.-H.}\ \bibnamefont {Deng}}, \bibinfo
  {author} {\bibfnamefont {M.-C.}\ \bibnamefont {Chen}}, \bibinfo {author}
  {\bibfnamefont {L.-C.}\ \bibnamefont {Peng}}, \bibinfo {author}
  {\bibfnamefont {Y.-H.}\ \bibnamefont {Luo}}, \bibinfo {author} {\bibfnamefont
  {J.}~\bibnamefont {Qin}}, \bibinfo {author} {\bibfnamefont {D.}~\bibnamefont
  {Wu}}, \bibinfo {author} {\bibfnamefont {X.}~\bibnamefont {Ding}}, \bibinfo
  {author} {\bibfnamefont {Y.}~\bibnamefont {Hu}}, \emph {et~al.},\ }\bibfield
  {title} {\bibinfo {title} {Quantum computational advantage using photons},\
  }\href@noop {} {\bibfield  {journal} {\bibinfo  {journal} {Sci.}\ }\textbf
  {\bibinfo {volume} {370}},\ \bibinfo {pages} {1460} (\bibinfo {year}
  {2020})}\BibitemShut {NoStop}%
\bibitem [{\citenamefont {Rivest}\ \emph {et~al.}(1978)\citenamefont {Rivest},
  \citenamefont {Shamir},\ and\ \citenamefont {Adleman}}]{rivest1978method}%
  \BibitemOpen
  \bibfield  {author} {\bibinfo {author} {\bibfnamefont {R.~L.}\ \bibnamefont
  {Rivest}}, \bibinfo {author} {\bibfnamefont {A.}~\bibnamefont {Shamir}},\
  and\ \bibinfo {author} {\bibfnamefont {L.}~\bibnamefont {Adleman}},\
  }\bibfield  {title} {\bibinfo {title} {A method for obtaining digital
  signatures and public-key cryptosystems},\ }\href@noop {} {\bibfield
  {journal} {\bibinfo  {journal} {Commun. ACM}\ }\textbf {\bibinfo {volume}
  {21}},\ \bibinfo {pages} {120} (\bibinfo {year} {1978})}\BibitemShut
  {NoStop}%
\bibitem [{\citenamefont {Shor}(1994)}]{shor1994algorithms}%
  \BibitemOpen
  \bibfield  {author} {\bibinfo {author} {\bibfnamefont {P.~W.}\ \bibnamefont
  {Shor}},\ }\bibfield  {title} {\bibinfo {title} {Algorithms for quantum
  computation: discrete logarithms and factoring},\ }in\ \href@noop {} {\emph
  {\bibinfo {booktitle} {Proceedings of 35th Annual Symposium on Foundations of
  Computer Science}}}\ (\bibinfo {publisher} {IEEE},\ \bibinfo {city} {New York, USA},\ \bibinfo {year} {1994})\
  pp.\ \bibinfo {pages} {124--134}\BibitemShut {NoStop}%
\bibitem [{\citenamefont {Shannon} (1949)}]{shannon1949communication}%
  \BibitemOpen
  \bibfield  {author} {\bibinfo {author} {\bibfnamefont {C.~E.}\ \bibnamefont
  {Shannon}},\ }\bibfield  {title} {\bibinfo {title} {Communication theory of
  secrecy systems},\ }\href@noop {} {\bibfield  {journal} {\bibinfo  {journal}
  {The Bell System Technical Journal}\ }\textbf {\bibinfo {volume} {28}},\
  \bibinfo {pages} {656} (\bibinfo {year} {1949})}\BibitemShut {NoStop}%
\bibitem [{\citenamefont {Bennett}\ and\ \citenamefont
  {Brassard}(1984)}]{bennett1984quantum}%
  \BibitemOpen
  \bibfield  {author} {\bibinfo {author} {\bibfnamefont {C.~H.}\ \bibnamefont
  {Bennett}}\ and\ \bibinfo {author} {\bibfnamefont {G.}~\bibnamefont
  {Brassard}},\ }\bibfield  {title} {\bibinfo {title} {Quantum cryptography:
  Public key distribution and coin tossing},\ }in\ \href@noop {} {\emph
  {\bibinfo {booktitle} {Proc. of the Int. Conf. on
  Comput., Syst. and Signal Process.}}}\ (\bibinfo {publisher} {IEEE},\ \bibinfo {city} {New York, USA},\ \bibinfo {year} {1984})\ pp.\ \bibinfo {pages} {175--179}\BibitemShut
  {NoStop}%
\bibitem [{\citenamefont {Ekert}(1991)}]{ekert1991quantum}%
  \BibitemOpen
  \bibfield  {author} {\bibinfo {author} {\bibfnamefont {A.~K.}\ \bibnamefont
  {Ekert}},\ }\bibfield  {title} {\bibinfo {title} {Quantum cryptography based on bell's theorem},\ }\href@noop {} {\bibfield  {journal} {\bibinfo
  {journal} {Phys. Rev. Lett.}\ }\textbf {\bibinfo {volume} {67}},\ \bibinfo
  {pages} {661} (\bibinfo {year} {1991})}\BibitemShut {NoStop}%
\bibitem [{\citenamefont {Weedbrook}\ \emph {et~al.}(2012)\citenamefont
  {Weedbrook}, \citenamefont {Pirandola}, \citenamefont {Garc\'{\i}a-Patr\'on},
  \citenamefont {Cerf}, \citenamefont {Ralph}, \citenamefont {Shapiro},\ and\
  \citenamefont {Lloyd}}]{RevModPhys.84.621}%
  \BibitemOpen
  \bibfield  {author} {\bibinfo {author} {\bibfnamefont {C.}~\bibnamefont
  {Weedbrook}}, \bibinfo {author} {\bibfnamefont {S.}~\bibnamefont
  {Pirandola}}, \bibinfo {author} {\bibfnamefont {R.}~\bibnamefont
  {Garc\'{\i}a-Patr\'on}}, \bibinfo {author} {\bibfnamefont {N.~J.}\
  \bibnamefont {Cerf}}, \bibinfo {author} {\bibfnamefont {T.~C.}\ \bibnamefont
  {Ralph}}, \bibinfo {author} {\bibfnamefont {J.~H.}\ \bibnamefont {Shapiro}},\
  and\ \bibinfo {author} {\bibfnamefont {S.}~\bibnamefont {Lloyd}},\ }\bibfield
   {title} {\bibinfo {title} {Gaussian quantum information},\ }\href@noop {}
  {\bibfield {journal} {\bibinfo  {journal} {Rev. Mod. Phys.}\ }\textbf
  {\bibinfo {volume} {84}},\ \bibinfo {pages} {621} (\bibinfo {year}
  {2012})}\BibitemShut {NoStop}%
\bibitem [{\citenamefont {Pirandola}\ \emph {et~al.}(2020)\citenamefont
  {Pirandola}, \citenamefont {Andersen}, \citenamefont {Banchi}, \citenamefont
  {Berta}, \citenamefont {Bunandar}, \citenamefont {Colbeck}, \citenamefont
  {Englund}, \citenamefont {Gehring}, \citenamefont {Lupo}, \citenamefont
  {Ottaviani} \emph {et~al.}}]{Pirandola2020advances}%
  \BibitemOpen
  \bibfield  {author} {\bibinfo {author} {\bibfnamefont {S.}~\bibnamefont
  {Pirandola}}, \bibinfo {author} {\bibfnamefont {U.~L.}\ \bibnamefont
  {Andersen}}, \bibinfo {author} {\bibfnamefont {L.}~\bibnamefont {Banchi}},
  \bibinfo {author} {\bibfnamefont {M.}~\bibnamefont {Berta}}, \bibinfo
  {author} {\bibfnamefont {D.}~\bibnamefont {Bunandar}}, \bibinfo {author}
  {\bibfnamefont {R.}~\bibnamefont {Colbeck}}, \bibinfo {author} {\bibfnamefont
  {D.}~\bibnamefont {Englund}}, \bibinfo {author} {\bibfnamefont
  {T.}~\bibnamefont {Gehring}}, \bibinfo {author} {\bibfnamefont
  {C.}~\bibnamefont {Lupo}}, \bibinfo {author} {\bibfnamefont {C.}~\bibnamefont
  {Ottaviani}}, \emph {et~al.},\ }\bibfield  {title} {\bibinfo {title}
  {Advances in quantum cryptography},\ }\href@noop {} {\bibfield  {journal}
  {\bibinfo {journal} {Adv. Opt. Photon.}\ }\textbf {\bibinfo {volume} {12}},\
  \bibinfo {pages} {1012} (\bibinfo {year} {2020})}\BibitemShut {NoStop}%
\bibitem [{\citenamefont {Xu}\ \emph {et~al.}(2020)\citenamefont {Xu},
  \citenamefont {Ma}, \citenamefont {Zhang}, \citenamefont {Lo},\ and\
  \citenamefont {Pan}}]{RevModPhys.92.025002}%
  \BibitemOpen
  \bibfield {author} {\bibinfo {author} {\bibfnamefont {F.}~\bibnamefont
  {Xu}}, \bibinfo {author} {\bibfnamefont {X.}~\bibnamefont {Ma}}, \bibinfo
  {author} {\bibfnamefont {Q.}~\bibnamefont {Zhang}}, \bibinfo {author}
  {\bibfnamefont {H.-K.}\ \bibnamefont {Lo}},\ and\ \bibinfo {author}
  {\bibfnamefont {J.-W.}\ \bibnamefont {Pan}},\ }\bibfield  {title} {\bibinfo
  {title} {Secure quantum key distribution with realistic devices},\ }\href
  {https://doi.org/10.1103/RevModPhys.92.025002} {\bibfield  {journal}
  {\bibinfo  {journal} {Rev. Mod. Phys.}\ }\textbf {\bibinfo {volume} {92}},\
  \bibinfo {pages} {025002} (\bibinfo {year} {2020})}\BibitemShut {NoStop}%
\bibitem [{\citenamefont {Zhao}\ \emph {et~al.}(2006)\citenamefont {Zhao},
  \citenamefont {Qi}, \citenamefont {Ma}, \citenamefont {Lo},\ and\
  \citenamefont {Qian}}]{PhysRevLett.96.070502}%
  \BibitemOpen
  \bibfield  {author} {\bibinfo {author} {\bibfnamefont {Y.}~\bibnamefont
  {Zhao}}, \bibinfo {author} {\bibfnamefont {B.}~\bibnamefont {Qi}}, \bibinfo
  {author} {\bibfnamefont {X.}~\bibnamefont {Ma}}, \bibinfo {author}
  {\bibfnamefont {H.-K.}\ \bibnamefont {Lo}},\ and\ \bibinfo {author}
  {\bibfnamefont {L.}~\bibnamefont {Qian}},\ }\bibfield  {title} {\bibinfo
  {title} {Experimental quantum key distribution with decoy states},\ }\href
  {https://doi.org/10.1103/PhysRevLett.96.070502} {\bibfield  {journal}
  {\bibinfo  {journal} {Phys. Rev. Lett.}\ }\textbf {\bibinfo {volume} {96}},\
  \bibinfo {pages} {070502} (\bibinfo {year} {2006})}\BibitemShut {NoStop}%
\bibitem [{\citenamefont {Comandar}\ \emph {et~al.}(2016)\citenamefont
  {Comandar}, \citenamefont {Lucamarini}, \citenamefont {Fr{\"o}hlich},
  \citenamefont {Dynes}, \citenamefont {Sharpe}, \citenamefont {Tam},
  \citenamefont {Yuan}, \citenamefont {Penty},\ and\ \citenamefont
  {Shields}}]{comandar2016quantum}%
  \BibitemOpen
  \bibfield  {author} {\bibinfo {author} {\bibfnamefont {L.~C.}\ \bibnamefont
  {Comandar}}, \bibinfo {author} {\bibfnamefont {M.}~\bibnamefont
  {Lucamarini}}, \bibinfo {author} {\bibfnamefont {B.}~\bibnamefont
  {Fr{\"o}hlich}}, \bibinfo {author} {\bibfnamefont {J.~F.}\ \bibnamefont
  {Dynes}}, \bibinfo {author} {\bibfnamefont {A.~W.}\ \bibnamefont {Sharpe}},
  \bibinfo {author} {\bibfnamefont {S.~W.~B.}\ \bibnamefont {Tam}}, \bibinfo
  {author} {\bibfnamefont {Z.~L.}\ \bibnamefont {Yuan}}, \bibinfo {author}
  {\bibfnamefont {R.~V.}\ \bibnamefont {Penty}},\ and\ \bibinfo {author}
  {\bibfnamefont {A.~J.}\ \bibnamefont {Shields}},\ }\bibfield  {title}
  {\bibinfo {title} {Quantum key distribution without detector vulnerabilities using optically seeded lasers},\ }\href@noop {} {\bibfield  {journal}
  {\bibinfo  {journal} {Nat. Photonics}\ }\textbf {\bibinfo {volume} {10}},\
  \bibinfo {pages} {312} (\bibinfo {year} {2016})}\BibitemShut {NoStop}%
\bibitem [{\citenamefont {Wei}\ \emph {et~al.}(2020)\citenamefont {Wei},
  \citenamefont {Li}, \citenamefont {Tan}, \citenamefont {Li}, \citenamefont
  {Min}, \citenamefont {Zhang}, \citenamefont {Li}, \citenamefont {You},
  \citenamefont {Wang}, \citenamefont {Jiang} \emph {et~al.}}]{wei2020high}%
  \BibitemOpen
  \bibfield  {author} {\bibinfo {author} {\bibfnamefont {K.}~\bibnamefont
  {Wei}}, \bibinfo {author} {\bibfnamefont {W.}~\bibnamefont {Li}}, \bibinfo
  {author} {\bibfnamefont {H.}~\bibnamefont {Tan}}, \bibinfo {author}
  {\bibfnamefont {Y.}~\bibnamefont {Li}}, \bibinfo {author} {\bibfnamefont
  {H.}~\bibnamefont {Min}}, \bibinfo {author} {\bibfnamefont {W.-J.}\
  \bibnamefont {Zhang}}, \bibinfo {author} {\bibfnamefont {H.}~\bibnamefont
  {Li}}, \bibinfo {author} {\bibfnamefont {L.}~\bibnamefont {You}}, \bibinfo
  {author} {\bibfnamefont {Z.}~\bibnamefont {Wang}}, \bibinfo {author}
  {\bibfnamefont {X.}~\bibnamefont {Jiang}}, \emph {et~al.},\ }\bibfield
  {title} {\bibinfo {title} {High-speed measurement-device-independent quantum key distribution with integrated silicon photonics},\ }\href
  {https://doi.org/10.1103/PhysRevX.10.031030} {\bibfield  {journal} {\bibinfo
  {journal} {Phys. Rev. X}\ }\textbf {\bibinfo {volume} {10}},\ \bibinfo
  {pages} {031030} (\bibinfo {year} {2020})}\BibitemShut {NoStop}%
\bibitem [{\citenamefont {Yin}\ \emph {et~al.}(2016)\citenamefont {Yin},
  \citenamefont {Chen}, \citenamefont {Yu}, \citenamefont {Liu}, \citenamefont
  {You}, \citenamefont {Zhou}, \citenamefont {Chen}, \citenamefont {Mao},
  \citenamefont {Huang}, \citenamefont {Zhang} \emph
  {et~al.}}]{yin2016measurement}%
  \BibitemOpen
  \bibfield  {author} {\bibinfo {author} {\bibfnamefont {H.-L.}\ \bibnamefont
  {Yin}}, \bibinfo {author} {\bibfnamefont {T.-Y.}\ \bibnamefont {Chen}},
  \bibinfo {author} {\bibfnamefont {Z.-W.}\ \bibnamefont {Yu}}, \bibinfo
  {author} {\bibfnamefont {H.}~\bibnamefont {Liu}}, \bibinfo {author}
  {\bibfnamefont {L.-X.}\ \bibnamefont {You}}, \bibinfo {author} {\bibfnamefont
  {Y.-H.}\ \bibnamefont {Zhou}}, \bibinfo {author} {\bibfnamefont {S.-J.}\
  \bibnamefont {Chen}}, \bibinfo {author} {\bibfnamefont {Y.}~\bibnamefont
  {Mao}}, \bibinfo {author} {\bibfnamefont {M.-Q.}\ \bibnamefont {Huang}},
  \bibinfo {author} {\bibfnamefont {W.-J.}\ \bibnamefont {Zhang}}, \emph
  {et~al.},\ }\bibfield  {title} {\bibinfo {title}
  {Measurement-device-independent quantum key distribution over a 404 km
  optical fiber},\ }\href {https://doi.org/10.1103/PhysRevLett.117.190501}
  {\bibfield  {journal} {\bibinfo  {journal} {Phys. Rev. Lett.}\ }\textbf
  {\bibinfo {volume} {117}},\ \bibinfo {pages} {190501} (\bibinfo {year}
  {2016})}\BibitemShut {NoStop}%
\bibitem [{\citenamefont {Boaron}\ \emph {et~al.}(2018)\citenamefont {Boaron},
  \citenamefont {Boso}, \citenamefont {Rusca}, \citenamefont {Vulliez},
  \citenamefont {Autebert}, \citenamefont {Caloz}, \citenamefont {Perrenoud},
  \citenamefont {Gras}, \citenamefont {Bussi\`eres}, \citenamefont {Li} \emph
  {et~al.}}]{boaron2018secure}%
  \BibitemOpen
  \bibfield  {author} {\bibinfo {author} {\bibfnamefont {A.}~\bibnamefont
  {Boaron}}, \bibinfo {author} {\bibfnamefont {G.}~\bibnamefont {Boso}},
  \bibinfo {author} {\bibfnamefont {D.}~\bibnamefont {Rusca}}, \bibinfo
  {author} {\bibfnamefont {C.}~\bibnamefont {Vulliez}}, \bibinfo {author}
  {\bibfnamefont {C.}~\bibnamefont {Autebert}}, \bibinfo {author}
  {\bibfnamefont {M.}~\bibnamefont {Caloz}}, \bibinfo {author} {\bibfnamefont
  {M.}~\bibnamefont {Perrenoud}}, \bibinfo {author} {\bibfnamefont
  {G.}~\bibnamefont {Gras}}, \bibinfo {author} {\bibfnamefont {F.}~\bibnamefont
  {Bussi\`eres}}, \bibinfo {author} {\bibfnamefont {M.-J.}\ \bibnamefont {Li}},
  \emph {et~al.},\ }\bibfield  {title} {\bibinfo {title} {Secure quantum key
  distribution over 421 km of optical fiber},\ }\href
  {https://doi.org/10.1103/PhysRevLett.121.190502} {\bibfield  {journal}
  {\bibinfo  {journal} {Phys. Rev. Lett.}\ }\textbf {\bibinfo {volume} {121}},\
  \bibinfo {pages} {190502} (\bibinfo {year} {2018})}\BibitemShut {NoStop}%
\bibitem [{\citenamefont {Yin}\ \emph {et~al.}(2020)\citenamefont {Yin},
  \citenamefont {Liu}, \citenamefont {Dai}, \citenamefont {Ci}, \citenamefont
  {Gu}, \citenamefont {Gao}, \citenamefont {Wang},\ and\ \citenamefont
  {Shen}}]{yin2020experimental}%
  \BibitemOpen
  \bibfield  {author} {\bibinfo {author} {\bibfnamefont {H.-L.}\ \bibnamefont
  {Yin}}, \bibinfo {author} {\bibfnamefont {P.}~\bibnamefont {Liu}}, \bibinfo
  {author} {\bibfnamefont {W.-W.}\ \bibnamefont {Dai}}, \bibinfo {author}
  {\bibfnamefont {Z.-H.}\ \bibnamefont {Ci}}, \bibinfo {author} {\bibfnamefont
  {J.}~\bibnamefont {Gu}}, \bibinfo {author} {\bibfnamefont {T.}~\bibnamefont
  {Gao}}, \bibinfo {author} {\bibfnamefont {Q.-W.}\ \bibnamefont {Wang}},\ and\
  \bibinfo {author} {\bibfnamefont {Z.-Y.}\ \bibnamefont {Shen}},\ }\bibfield
  {title} {\bibinfo {title} {Experimental composable security decoy-state
  quantum key distribution using time-phase encoding},\ }\href@noop {}
  {\bibfield {journal} {\bibinfo  {journal} {Opt. Express}\ }\textbf {\bibinfo
  {volume} {28}},\ \bibinfo {pages} {29479} (\bibinfo {year}
  {2020})}\BibitemShut {NoStop}%
\bibitem [{\citenamefont {Joshi}\ \emph {et~al.}(2020)\citenamefont {Joshi},
  \citenamefont {Aktas}, \citenamefont {Wengerowsky}, \citenamefont {Lon{\v
  c}ari{\'c}}, \citenamefont {Neumann}, \citenamefont {Liu}, \citenamefont
  {Scheidl}, \citenamefont {Lorenzo}, \citenamefont {Samec}, \citenamefont
  {Kling} \emph {et~al.}}]{Joshieaba0959}%
  \BibitemOpen
  \bibfield  {author} {\bibinfo {author} {\bibfnamefont {S.~K.}\ \bibnamefont
  {Joshi}}, \bibinfo {author} {\bibfnamefont {D.}~\bibnamefont {Aktas}},
  \bibinfo {author} {\bibfnamefont {S.}~\bibnamefont {Wengerowsky}}, \bibinfo
  {author} {\bibfnamefont {M.}~\bibnamefont {Lon{\v c}ari{\'c}}}, \bibinfo
  {author} {\bibfnamefont {S.~P.}\ \bibnamefont {Neumann}}, \bibinfo {author}
  {\bibfnamefont {B.}~\bibnamefont {Liu}}, \bibinfo {author} {\bibfnamefont
  {T.}~\bibnamefont {Scheidl}}, \bibinfo {author} {\bibfnamefont {G.~C.}\
  \bibnamefont {Lorenzo}}, \bibinfo {author} {\bibfnamefont {{\v
  Z}.}~\bibnamefont {Samec}}, \bibinfo {author} {\bibfnamefont
  {L.}~\bibnamefont {Kling}}, \emph {et~al.},\ }\bibfield  {title} {\bibinfo
  {title} {A trusted node{\textendash}free eight-user metropolitan quantum
  communication network},\ }\href@noop {} {\bibfield  {journal} {\bibinfo
  {journal} {Sci. Adv.}\ }\textbf {\bibinfo {volume} {6}},\ \bibinfo {pages}
  {eaba0959} (\bibinfo {year} {2020})}\BibitemShut {NoStop}%
\bibitem [{\citenamefont {Tang}\ \emph {et~al.}(2016)\citenamefont {Tang},
  \citenamefont {Yin}, \citenamefont {Zhao}, \citenamefont {Liu}, \citenamefont
  {Sun}, \citenamefont {Huang}, \citenamefont {Zhang}, \citenamefont {Chen},
  \citenamefont {Zhang}, \citenamefont {You} \emph
  {et~al.}}]{tang2016measurement}%
  \BibitemOpen
  \bibfield  {author} {\bibinfo {author} {\bibfnamefont {Y.-L.}\ \bibnamefont
  {Tang}}, \bibinfo {author} {\bibfnamefont {H.-L.}\ \bibnamefont {Yin}},
  \bibinfo {author} {\bibfnamefont {Q.}~\bibnamefont {Zhao}}, \bibinfo {author}
  {\bibfnamefont {H.}~\bibnamefont {Liu}}, \bibinfo {author} {\bibfnamefont
  {X.-X.}\ \bibnamefont {Sun}}, \bibinfo {author} {\bibfnamefont {M.-Q.}\
  \bibnamefont {Huang}}, \bibinfo {author} {\bibfnamefont {W.-J.}\ \bibnamefont
  {Zhang}}, \bibinfo {author} {\bibfnamefont {S.-J.}\ \bibnamefont {Chen}},
  \bibinfo {author} {\bibfnamefont {L.}~\bibnamefont {Zhang}}, \bibinfo
  {author} {\bibfnamefont {L.-X.}\ \bibnamefont {You}}, \emph {et~al.},\
  }\bibfield  {title} {\bibinfo {title} {Measurement-device-independent quantum
  key distribution over untrustful metropolitan network},\ }\href
  {https://doi.org/10.1103/PhysRevX.6.011024} {\bibfield  {journal} {\bibinfo
  {journal} {Phys. Rev. X}\ }\textbf {\bibinfo {volume} {6}},\ \bibinfo {pages}
  {011024} (\bibinfo {year} {2016})}\BibitemShut {NoStop}%
\bibitem [{\citenamefont {Diamanti}\ and\ \citenamefont
  {Leverrier}(2015)}]{diamanti2015distributing}%
  \BibitemOpen
  \bibfield  {author} {\bibinfo {author} {\bibfnamefont {E.}~\bibnamefont
  {Diamanti}}\ and\ \bibinfo {author} {\bibfnamefont {A.}~\bibnamefont
  {Leverrier}},\ }\bibfield  {title} {\bibinfo {title} {Distributing secret
  keys with quantum continuous variables: Principle, security and
  implementations},\ }\href {https://doi.org/10.3390/e17096072} {\bibfield
  {journal} {\bibinfo  {journal} {Entropy}\ }\textbf {\bibinfo {volume} {17}},\
  \bibinfo {pages} {6072} (\bibinfo {year} {2015})}\BibitemShut {NoStop}%
\bibitem [{\citenamefont {Grosshans}\ and\ \citenamefont
  {Grangier}(2002)}]{grosshans2002continuous}%
  \BibitemOpen
  \bibfield  {author} {\bibinfo {author} {\bibfnamefont {F.}~\bibnamefont
  {Grosshans}}\ and\ \bibinfo {author} {\bibfnamefont {P.}~\bibnamefont
  {Grangier}},\ }\bibfield  {title} {\bibinfo {title} {Continuous variable
  quantum cryptography using coherent states},\ }\href@noop {} {\bibfield
  {journal} {\bibinfo  {journal} {Phys. Rev. Lett.}\ }\textbf {\bibinfo
  {volume} {88}},\ \bibinfo {pages} {057902} (\bibinfo {year}
  {2002})}\BibitemShut {NoStop}%
\bibitem [{\citenamefont {Grosshans}\ \emph {et~al.}(2003)\citenamefont
  {Grosshans}, \citenamefont {Van~Assche}, \citenamefont {Wenger},
  \citenamefont {Brouri}, \citenamefont {Cerf},\ and\ \citenamefont
  {Grangier}}]{grosshans2003quantum}%
  \BibitemOpen
  \bibfield  {author} {\bibinfo {author} {\bibfnamefont {F.}~\bibnamefont
  {Grosshans}}, \bibinfo {author} {\bibfnamefont {G.}~\bibnamefont
  {Van~Assche}}, \bibinfo {author} {\bibfnamefont {J.}~\bibnamefont {Wenger}},
  \bibinfo {author} {\bibfnamefont {R.}~\bibnamefont {Brouri}}, \bibinfo
  {author} {\bibfnamefont {N.~J.}\ \bibnamefont {Cerf}},\ and\ \bibinfo
  {author} {\bibfnamefont {P.}~\bibnamefont {Grangier}},\ }\bibfield  {title}
  {\bibinfo {title} {Quantum key distribution using gaussian-modulated coherent
  states},\ }\href@noop {} {\bibfield  {journal} {\bibinfo  {journal} {Nature}\
  }\textbf {\bibinfo {volume} {421}},\ \bibinfo {pages} {238} (\bibinfo {year}
  {2003})}\BibitemShut {NoStop}%
\bibitem [{\citenamefont {Yin}\ \emph {et~al.}(2019)\citenamefont {Yin},
  \citenamefont {Zhu},\ and\ \citenamefont {Fu}}]{yin2019phase}%
  \BibitemOpen
  \bibfield  {author} {\bibinfo {author} {\bibfnamefont {H.-L.}\ \bibnamefont
  {Yin}}, \bibinfo {author} {\bibfnamefont {W.}~\bibnamefont {Zhu}},\ and\
  \bibinfo {author} {\bibfnamefont {Y.}~\bibnamefont {Fu}},\ }\bibfield
  {title} {\bibinfo {title} {Phase self-aligned continuous-variable
  measurement-device-independent quantum key distribution},\ }\href@noop {}
  {\bibfield  {journal} {\bibinfo  {journal} {Sci. Rep.}\ }\textbf {\bibinfo
  {volume} {9}},\ \bibinfo {pages} {49} (\bibinfo {year} {2019})}\BibitemShut
  {NoStop}%
\bibitem [{\citenamefont {Cerf}\ \emph {et~al.}(2001)\citenamefont {Cerf},
  \citenamefont {Levy},\ and\ \citenamefont {Van~Assche}}]{cerf2001quantum}%
  \BibitemOpen
  \bibfield  {author} {\bibinfo {author} {\bibfnamefont {N.~J.}\ \bibnamefont
  {Cerf}}, \bibinfo {author} {\bibfnamefont {M.}~\bibnamefont {Levy}},\ and\
  \bibinfo {author} {\bibfnamefont {G.}~\bibnamefont {Van~Assche}},\ }\bibfield
   {title} {\bibinfo {title} {Quantum distribution of gaussian keys using
  squeezed states},\ }\href@noop {} {\bibfield  {journal} {\bibinfo  {journal}
  {Phys. Rev. A}\ }\textbf {\bibinfo {volume} {63}},\ \bibinfo {pages} {052311}
  (\bibinfo {year} {2001})}\BibitemShut {NoStop}%
\bibitem [{\citenamefont {Lodewyck}\ \emph {et~al.}(2007)\citenamefont
  {Lodewyck}, \citenamefont {Bloch}, \citenamefont {Garc\'{\i}a-Patr\'on},
  \citenamefont {Fossier}, \citenamefont {Karpov}, \citenamefont {Diamanti},
  \citenamefont {Debuisschert}, \citenamefont {Cerf}, \citenamefont
  {Tualle-Brouri}, \citenamefont {McLaughlin},\ and\ \citenamefont
  {Grangier}}]{lodewyck2007quantum}%
  \BibitemOpen
  \bibfield  {author} {\bibinfo {author} {\bibfnamefont {J.}~\bibnamefont
  {Lodewyck}}, \bibinfo {author} {\bibfnamefont {M.}~\bibnamefont {Bloch}},
  \bibinfo {author} {\bibfnamefont {R.}~\bibnamefont {Garc\'{\i}a-Patr\'on}},
  \bibinfo {author} {\bibfnamefont {S.}~\bibnamefont {Fossier}}, \bibinfo
  {author} {\bibfnamefont {E.}~\bibnamefont {Karpov}}, \bibinfo {author}
  {\bibfnamefont {E.}~\bibnamefont {Diamanti}}, \bibinfo {author}
  {\bibfnamefont {T.}~\bibnamefont {Debuisschert}}, \bibinfo {author}
  {\bibfnamefont {N.~J.}\ \bibnamefont {Cerf}}, \bibinfo {author}
  {\bibfnamefont {R.}~\bibnamefont {Tualle-Brouri}}, \bibinfo {author}
  {\bibfnamefont {S.~W.}\ \bibnamefont {McLaughlin}},\ and\ \bibinfo {author}
  {\bibfnamefont {P.}~\bibnamefont {Grangier}},\ }\bibfield  {title} {\bibinfo
  {title} {Quantum key distribution over 25 km with an all-fiber
  continuous-variable system},\ }\href
  {https://doi.org/10.1103/PhysRevA.76.042305} {\bibfield  {journal} {\bibinfo
  {journal} {Phys. Rev. A}\ }\textbf {\bibinfo {volume} {76}},\ \bibinfo
  {pages} {042305} (\bibinfo {year} {2007})}\BibitemShut {NoStop}%
\bibitem [{\citenamefont {Qi}\ \emph {et~al.}(2007)\citenamefont {Qi},
  \citenamefont {Huang}, \citenamefont {Qian},\ and\ \citenamefont
  {Lo}}]{qi2007experimental}%
  \BibitemOpen
  \bibfield  {author} {\bibinfo {author} {\bibfnamefont {B.}~\bibnamefont
  {Qi}}, \bibinfo {author} {\bibfnamefont {L.-L.}\ \bibnamefont {Huang}},
  \bibinfo {author} {\bibfnamefont {L.}~\bibnamefont {Qian}},\ and\ \bibinfo
  {author} {\bibfnamefont {H.-K.}\ \bibnamefont {Lo}},\ }\bibfield  {title}
  {\bibinfo {title} {Experimental study on the gaussian-modulated
  coherent-state quantum key distribution over standard telecommunication
  fibers},\ }\href {https://doi.org/10.1103/PhysRevA.76.052323} {\bibfield
  {journal} {\bibinfo  {journal} {Phys. Rev. A}\ }\textbf {\bibinfo {volume}
  {76}},\ \bibinfo {pages} {052323} (\bibinfo {year} {2007})}\BibitemShut
  {NoStop}%
\bibitem [{\citenamefont {Fossier}\ \emph {et~al.}(2009)\citenamefont
  {Fossier}, \citenamefont {Diamanti}, \citenamefont {Debuisschert},
  \citenamefont {Villing}, \citenamefont {Tualle-Brouri},\ and\ \citenamefont
  {Grangier}}]{fossier2009field}%
  \BibitemOpen
  \bibfield  {author} {\bibinfo {author} {\bibfnamefont {S.}~\bibnamefont
  {Fossier}}, \bibinfo {author} {\bibfnamefont {E.}~\bibnamefont {Diamanti}},
  \bibinfo {author} {\bibfnamefont {T.}~\bibnamefont {Debuisschert}}, \bibinfo
  {author} {\bibfnamefont {A.}~\bibnamefont {Villing}}, \bibinfo {author}
  {\bibfnamefont {R.}~\bibnamefont {Tualle-Brouri}},\ and\ \bibinfo {author}
  {\bibfnamefont {P.}~\bibnamefont {Grangier}},\ }\bibfield  {title} {\bibinfo
  {title} {Field test of a continuous-variable quantum key distribution
  prototype},\ }\href@noop {} {\bibfield  {journal} {\bibinfo  {journal} {New
  J. Phys.}\ }\textbf {\bibinfo {volume} {11}},\ \bibinfo {pages} {045023}
  (\bibinfo {year} {2009})}\BibitemShut {NoStop}%
\bibitem [{\citenamefont {Jouguet}\ \emph {et~al.}(2013)\citenamefont
  {Jouguet}, \citenamefont {Kunz-Jacques}, \citenamefont {Leverrier},
  \citenamefont {Grangier},\ and\ \citenamefont
  {Diamanti}}]{jouguet2013experimental}%
  \BibitemOpen
  \bibfield  {author} {\bibinfo {author} {\bibfnamefont {P.}~\bibnamefont
  {Jouguet}}, \bibinfo {author} {\bibfnamefont {S.}~\bibnamefont
  {Kunz-Jacques}}, \bibinfo {author} {\bibfnamefont {A.}~\bibnamefont
  {Leverrier}}, \bibinfo {author} {\bibfnamefont {P.}~\bibnamefont
  {Grangier}},\ and\ \bibinfo {author} {\bibfnamefont {E.}~\bibnamefont
  {Diamanti}},\ }\bibfield  {title} {\bibinfo {title} {Experimental
  demonstration of long-distance continuous-variable quantum key
  distribution},\ }\href@noop {} {\bibfield  {journal} {\bibinfo  {journal}
  {Nat. Photonics}\ }\textbf {\bibinfo {volume} {7}},\ \bibinfo {pages} {378}
  (\bibinfo {year} {2013})}\BibitemShut {NoStop}%
\bibitem [{\citenamefont {Huang}\ \emph {et~al.}(2016)\citenamefont {Huang},
  \citenamefont {Huang}, \citenamefont {Lin},\ and\ \citenamefont
  {Zeng}}]{huang2016long}%
  \BibitemOpen
  \bibfield  {author} {\bibinfo {author} {\bibfnamefont {D.}~\bibnamefont
  {Huang}}, \bibinfo {author} {\bibfnamefont {P.}~\bibnamefont {Huang}},
  \bibinfo {author} {\bibfnamefont {D.}~\bibnamefont {Lin}},\ and\ \bibinfo
  {author} {\bibfnamefont {G.}~\bibnamefont {Zeng}},\ }\bibfield  {title}
  {\bibinfo {title} {Long-distance continuous-variable quantum key distribution by controlling excess noise},\ }\href@noop {} {\bibfield  {journal} {\bibinfo
   {journal} {Sci. Rep.}\ }\textbf {\bibinfo {volume} {6}},\ \bibinfo {pages}
  {19201} (\bibinfo {year} {2016})}\BibitemShut {NoStop}%
\bibitem [{\citenamefont {Liu}\ \emph {et~al.}(2020)\citenamefont {Liu},
  \citenamefont {Cao}, \citenamefont {Wang},\ and\ \citenamefont
  {Li}}]{PhysRevA.102.032625}%
  \BibitemOpen
  \bibfield  {author} {\bibinfo {author} {\bibfnamefont {W.}~\bibnamefont
  {Liu}}, \bibinfo {author} {\bibfnamefont {Y.}~\bibnamefont {Cao}}, \bibinfo
  {author} {\bibfnamefont {X.}~\bibnamefont {Wang}},\ and\ \bibinfo {author}
  {\bibfnamefont {Y.}~\bibnamefont {Li}},\ }\bibfield  {title} {\bibinfo
  {title} {Continuous-variable quantum key distribution under strong channel
  polarization disturbance},\ }\href
  {https://doi.org/10.1103/PhysRevA.102.032625} {\bibfield  {journal} {\bibinfo
   {journal} {Phys. Rev. A}\ }\textbf {\bibinfo {volume} {102}},\ \bibinfo
  {pages} {032625} (\bibinfo {year} {2020})}\BibitemShut {NoStop}%
\bibitem [{\citenamefont {Qi}\ \emph {et~al.}(2015)\citenamefont {Qi},
  \citenamefont {Lougovski}, \citenamefont {Pooser}, \citenamefont {Grice},\
  and\ \citenamefont {Bobrek}}]{PhysRevX.5.041009}%
  \BibitemOpen
  \bibfield  {author} {\bibinfo {author} {\bibfnamefont {B.}~\bibnamefont
  {Qi}}, \bibinfo {author} {\bibfnamefont {P.}~\bibnamefont {Lougovski}},
  \bibinfo {author} {\bibfnamefont {R.}~\bibnamefont {Pooser}}, \bibinfo
  {author} {\bibfnamefont {W.}~\bibnamefont {Grice}},\ and\ \bibinfo {author}
  {\bibfnamefont {M.}~\bibnamefont {Bobrek}},\ }\bibfield  {title} {\bibinfo
  {title} {Generating the local oscillator ``locally'' in continuous-variable
  quantum key distribution based on coherent detection},\ }\href
  {https://doi.org/10.1103/PhysRevX.5.041009} {\bibfield  {journal} {\bibinfo
  {journal} {Phys. Rev. X}\ }\textbf {\bibinfo {volume} {5}},\ \bibinfo {pages}
  {041009} (\bibinfo {year} {2015})}\BibitemShut {NoStop}%
\bibitem [{\citenamefont {Soh}\ \emph {et~al.}(2015)\citenamefont {Soh},
  \citenamefont {Brif}, \citenamefont {Coles}, \citenamefont {L\"utkenhaus},
  \citenamefont {Camacho}, \citenamefont {Urayama},\ and\ \citenamefont
  {Sarovar}}]{PhysRevX.5.041010}%
  \BibitemOpen
  \bibfield  {author} {\bibinfo {author} {\bibfnamefont {D.~B.~S.}\
  \bibnamefont {Soh}}, \bibinfo {author} {\bibfnamefont {C.}~\bibnamefont
  {Brif}}, \bibinfo {author} {\bibfnamefont {P.~J.}\ \bibnamefont {Coles}},
  \bibinfo {author} {\bibfnamefont {N.}~\bibnamefont {L\"utkenhaus}}, \bibinfo
  {author} {\bibfnamefont {R.~M.}\ \bibnamefont {Camacho}}, \bibinfo {author}
  {\bibfnamefont {J.}~\bibnamefont {Urayama}},\ and\ \bibinfo {author}
  {\bibfnamefont {M.}~\bibnamefont {Sarovar}},\ }\bibfield  {title} {\bibinfo
  {title} {Self-referenced continuous-variable quantum key distribution
  protocol},\ }\href {https://doi.org/10.1103/PhysRevX.5.041010} {\bibfield
  {journal} {\bibinfo  {journal} {Phys. Rev. X}\ }\textbf {\bibinfo {volume}
  {5}},\ \bibinfo {pages} {041010} (\bibinfo {year} {2015})}\BibitemShut
  {NoStop}%
\bibitem [{\citenamefont {Huang}\ \emph {et~al.}(2015)\citenamefont {Huang},
  \citenamefont {Huang}, \citenamefont {Lin}, \citenamefont {Wang},\ and\
  \citenamefont {Zeng}}]{huang2015high}%
  \BibitemOpen
  \bibfield  {author} {\bibinfo {author} {\bibfnamefont {D.}~\bibnamefont
  {Huang}}, \bibinfo {author} {\bibfnamefont {P.}~\bibnamefont {Huang}},
  \bibinfo {author} {\bibfnamefont {D.}~\bibnamefont {Lin}}, \bibinfo {author}
  {\bibfnamefont {C.}~\bibnamefont {Wang}},\ and\ \bibinfo {author}
  {\bibfnamefont {G.}~\bibnamefont {Zeng}},\ }\bibfield  {title} {\bibinfo
  {title} {High-speed continuous-variable quantum key distribution without
  sending a local oscillator},\ }\href {https://doi.org/10.1364/OL.40.003695}
  {\bibfield  {journal} {\bibinfo  {journal} {Opt. Lett.}\ }\textbf {\bibinfo
  {volume} {40}},\ \bibinfo {pages} {3695} (\bibinfo {year}
  {2015})}\BibitemShut {NoStop}%
\bibitem [{\citenamefont {Pirandola}\ \emph {et~al.}(2015)\citenamefont
  {Pirandola}, \citenamefont {Ottaviani}, \citenamefont {Spedalieri},
  \citenamefont {Weedbrook}, \citenamefont {Braunstein}, \citenamefont {Lloyd},
  \citenamefont {Gehring}, \citenamefont {Jacobsen},\ and\ \citenamefont
  {Andersen}}]{pirandola2015high}%
  \BibitemOpen
  \bibfield  {author} {\bibinfo {author} {\bibfnamefont {S.}~\bibnamefont
  {Pirandola}}, \bibinfo {author} {\bibfnamefont {C.}~\bibnamefont
  {Ottaviani}}, \bibinfo {author} {\bibfnamefont {G.}~\bibnamefont
  {Spedalieri}}, \bibinfo {author} {\bibfnamefont {C.}~\bibnamefont
  {Weedbrook}}, \bibinfo {author} {\bibfnamefont {S.~L.}\ \bibnamefont
  {Braunstein}}, \bibinfo {author} {\bibfnamefont {S.}~\bibnamefont {Lloyd}},
  \bibinfo {author} {\bibfnamefont {T.}~\bibnamefont {Gehring}}, \bibinfo
  {author} {\bibfnamefont {C.~S.}\ \bibnamefont {Jacobsen}},\ and\ \bibinfo
  {author} {\bibfnamefont {U.~L.}\ \bibnamefont {Andersen}},\ }\bibfield
  {title} {\bibinfo {title} {High-rate measurement-device-independent quantum
  cryptography},\ }\href@noop {} {\bibfield  {journal} {\bibinfo  {journal}
  {Nat. Photonics}\ }\textbf {\bibinfo {volume} {9}},\ \bibinfo {pages} {397}
  (\bibinfo {year} {2015})}\BibitemShut {NoStop}%
\bibitem [{\citenamefont {Zhang}\ \emph {et~al.}(2019)\citenamefont {Zhang},
  \citenamefont {Haw}, \citenamefont {Cai}, \citenamefont {Xu}, \citenamefont
  {Assad}, \citenamefont {Fitzsimons}, \citenamefont {Zhou}, \citenamefont
  {Zhang}, \citenamefont {Yu}, \citenamefont {Wu} \emph
  {et~al.}}]{zhang2019integrated}%
  \BibitemOpen
  \bibfield  {author} {\bibinfo {author} {\bibfnamefont {G.}~\bibnamefont
  {Zhang}}, \bibinfo {author} {\bibfnamefont {J.~Y.}\ \bibnamefont {Haw}},
  \bibinfo {author} {\bibfnamefont {H.}~\bibnamefont {Cai}}, \bibinfo {author}
  {\bibfnamefont {F.}~\bibnamefont {Xu}}, \bibinfo {author} {\bibfnamefont
  {S.}~\bibnamefont {Assad}}, \bibinfo {author} {\bibfnamefont {J.~F.}\
  \bibnamefont {Fitzsimons}}, \bibinfo {author} {\bibfnamefont
  {X.}~\bibnamefont {Zhou}}, \bibinfo {author} {\bibfnamefont {Y.}~\bibnamefont
  {Zhang}}, \bibinfo {author} {\bibfnamefont {S.}~\bibnamefont {Yu}}, \bibinfo
  {author} {\bibfnamefont {J.}~\bibnamefont {Wu}}, \emph {et~al.},\ }\bibfield
  {title} {\bibinfo {title} {An integrated silicon photonic chip platform for
  continuous-variable quantum key distribution},\ }\href@noop {} {\bibfield
  {journal} {\bibinfo  {journal} {Nat. Photonics}\ }\textbf {\bibinfo {volume}
  {13}},\ \bibinfo {pages} {839} (\bibinfo {year} {2019})}\BibitemShut
  {NoStop}%
\bibitem [{\citenamefont {Zhang}\ \emph {et~al.}(2020)\citenamefont {Zhang},
  \citenamefont {Chen}, \citenamefont {Pirandola}, \citenamefont {Wang},
  \citenamefont {Zhou}, \citenamefont {Chu}, \citenamefont {Zhao},
  \citenamefont {Xu}, \citenamefont {Yu},\ and\ \citenamefont
  {Guo}}]{zhang2020long}%
  \BibitemOpen
  \bibfield  {author} {\bibinfo {author} {\bibfnamefont {Y.}~\bibnamefont
  {Zhang}}, \bibinfo {author} {\bibfnamefont {Z.}~\bibnamefont {Chen}},
  \bibinfo {author} {\bibfnamefont {S.}~\bibnamefont {Pirandola}}, \bibinfo
  {author} {\bibfnamefont {X.}~\bibnamefont {Wang}}, \bibinfo {author}
  {\bibfnamefont {C.}~\bibnamefont {Zhou}}, \bibinfo {author} {\bibfnamefont
  {B.}~\bibnamefont {Chu}}, \bibinfo {author} {\bibfnamefont {Y.}~\bibnamefont
  {Zhao}}, \bibinfo {author} {\bibfnamefont {B.}~\bibnamefont {Xu}}, \bibinfo
  {author} {\bibfnamefont {S.}~\bibnamefont {Yu}},\ and\ \bibinfo {author}
  {\bibfnamefont {H.}~\bibnamefont {Guo}},\ }\bibfield  {title} {\bibinfo
  {title} {Long-distance continuous-variable quantum key distribution over
  202.81 km fiber},\ }\href@noop {} {\bibfield  {journal} {\bibinfo  {journal}
  {Phys. Rev. Lett.}\ }\textbf {\bibinfo {volume} {125}},\ \bibinfo {pages}
  {010502} (\bibinfo {year} {2020})}\BibitemShut {NoStop}%
\bibitem [{\citenamefont {Silberhorn}\ \emph {et~al.}(2002)\citenamefont
  {Silberhorn}, \citenamefont {Ralph}, \citenamefont {L{\"u}tkenhaus},\ and\
  \citenamefont {Leuchs}}]{silberhorn2002continuous}%
  \BibitemOpen
  \bibfield  {author} {\bibinfo {author} {\bibfnamefont {C.}~\bibnamefont
  {Silberhorn}}, \bibinfo {author} {\bibfnamefont {T.~C.}\ \bibnamefont
  {Ralph}}, \bibinfo {author} {\bibfnamefont {N.}~\bibnamefont
  {L{\"u}tkenhaus}},\ and\ \bibinfo {author} {\bibfnamefont {G.}~\bibnamefont
  {Leuchs}},\ }\bibfield  {title} {\bibinfo {title} {Continuous variable
  quantum cryptography: Beating the 3 db loss limit},\ }\href@noop {}
  {\bibfield  {journal} {\bibinfo  {journal} {Phys. Rev. Lett.}\ }\textbf
  {\bibinfo {volume} {89}},\ \bibinfo {pages} {167901} (\bibinfo {year}
  {2002})}\BibitemShut {NoStop}%
\bibitem [{\citenamefont {Garc{\'\i}a-Patr{\'o}n}\ and\ \citenamefont
  {Cerf}(2009)}]{garcia2009continuous}%
  \BibitemOpen
  \bibfield  {author} {\bibinfo {author} {\bibfnamefont {R.}~\bibnamefont
  {Garc{\'\i}a-Patr{\'o}n}}\ and\ \bibinfo {author} {\bibfnamefont {N.~J.}\
  \bibnamefont {Cerf}},\ }\bibfield  {title} {\bibinfo {title}
  {Continuous-variable quantum key distribution protocols over noisy
  channels},\ }\href@noop {} {\bibfield  {journal} {\bibinfo  {journal} {Phys.
  Rev. Lett.}\ }\textbf {\bibinfo {volume} {102}},\ \bibinfo {pages} {130501}
  (\bibinfo {year} {2009})}\BibitemShut {NoStop}%
\bibitem [{\citenamefont {Leverrier}\ \emph {et~al.}(2010)\citenamefont
  {Leverrier}, \citenamefont {Grosshans},\ and\ \citenamefont
  {Grangier}}]{leverrier2010finite}%
  \BibitemOpen
  \bibfield  {author} {\bibinfo {author} {\bibfnamefont {A.}~\bibnamefont
  {Leverrier}}, \bibinfo {author} {\bibfnamefont {F.}~\bibnamefont
  {Grosshans}},\ and\ \bibinfo {author} {\bibfnamefont {P.}~\bibnamefont
  {Grangier}},\ }\bibfield  {title} {\bibinfo {title} {Finite-size analysis of a continuous-variable quantum key distribution},\ }\href
  {https://doi.org/10.1103/PhysRevA.81.062343} {\bibfield  {journal} {\bibinfo
  {journal} {Phys. Rev. A}\ }\textbf {\bibinfo {volume} {81}},\ \bibinfo
  {pages} {062343} (\bibinfo {year} {2010})}\BibitemShut {NoStop}%
\bibitem [{\citenamefont {Leverrier}(2015)}]{leverrier2015composable}%
  \BibitemOpen
  \bibfield  {author} {\bibinfo {author} {\bibfnamefont {A.}~\bibnamefont
  {Leverrier}},\ }\bibfield  {title} {\bibinfo {title} {Composable security
  proof for continuous-variable quantum key distribution with coherent
  states},\ }\href {https://doi.org/10.1103/PhysRevLett.114.070501} {\bibfield
  {journal} {\bibinfo  {journal} {Phys. Rev. Lett.}\ }\textbf {\bibinfo
  {volume} {114}},\ \bibinfo {pages} {070501} (\bibinfo {year}
  {2015})}\BibitemShut {NoStop}%
\bibitem [{\citenamefont {Pirandola}(2021)}]{PhysRevResearch.3.013279}%
  \BibitemOpen
  \bibfield  {author} {\bibinfo {author} {\bibfnamefont {S.}~\bibnamefont
  {Pirandola}},\ }\bibfield  {title} {\bibinfo {title} {Limits and security of free-space quantum communications},\ }\href
  {https://doi.org/10.1103/PhysRevResearch.3.013279} {\bibfield  {journal}
  {\bibinfo  {journal} {Phys. Rev. Research}\ }\textbf {\bibinfo {volume}
  {3}},\ \bibinfo {pages} {013279} (\bibinfo {year} {2021})}\BibitemShut
  {NoStop}%
\bibitem [{\citenamefont {Navascu\'es}\ \emph {et~al.}(2006)\citenamefont
  {Navascu\'es}, \citenamefont {Grosshans},\ and\ \citenamefont
  {Ac\'{\i}n}}]{PhysRevLett.97.190502}%
  \BibitemOpen
  \bibfield  {author} {\bibinfo {author} {\bibfnamefont {M.}~\bibnamefont
  {Navascu\'es}}, \bibinfo {author} {\bibfnamefont {F.}~\bibnamefont
  {Grosshans}},\ and\ \bibinfo {author} {\bibfnamefont {A.}~\bibnamefont
  {Ac\'{\i}n}},\ }\bibfield  {title} {\bibinfo {title} {Optimality of gaussian
  attacks in continuous-variable quantum cryptography},\ }\href
  {https://doi.org/10.1103/PhysRevLett.97.190502} {\bibfield  {journal}
  {\bibinfo  {journal} {Phys. Rev. Lett.}\ }\textbf {\bibinfo {volume} {97}},\
  \bibinfo {pages} {190502} (\bibinfo {year} {2006})}\BibitemShut {NoStop}%
\bibitem [{\citenamefont {Garc\'{\i}a-Patr\'on}\ and\ \citenamefont
  {Cerf}(2006)}]{PhysRevLett.97.190503}%
  \BibitemOpen
  \bibfield  {author} {\bibinfo {author} {\bibfnamefont {R.}~\bibnamefont
  {Garc\'{\i}a-Patr\'on}}\ and\ \bibinfo {author} {\bibfnamefont {N.~J.}\
  \bibnamefont {Cerf}},\ }\bibfield  {title} {\bibinfo {title} {Unconditional
  optimality of gaussian attacks against continuous-variable quantum key
  distribution},\ }\href {https://doi.org/10.1103/PhysRevLett.97.190503}
  {\bibfield  {journal} {\bibinfo  {journal} {Phys. Rev. Lett.}\ }\textbf
  {\bibinfo {volume} {97}},\ \bibinfo {pages} {190503} (\bibinfo {year}
  {2006})}\BibitemShut {NoStop}%
\bibitem [{\citenamefont {Lupo}(2020)}]{lupo2020towards}%
  \BibitemOpen
  \bibfield  {author} {\bibinfo {author} {\bibfnamefont {C.}~\bibnamefont
  {Lupo}},\ }\bibfield  {title} {\bibinfo {title} {Towards practical security
  of continuous-variable quantum key distribution},\ }\href
  {https://doi.org/10.1103/PhysRevA.102.022623} {\bibfield  {journal} {\bibinfo
   {journal} {Phys. Rev. A}\ }\textbf {\bibinfo {volume} {102}},\ \bibinfo
  {pages} {022623} (\bibinfo {year} {2020})}\BibitemShut {NoStop}%
\bibitem [{\citenamefont {Zhou}\ \emph {et~al.}(2019)\citenamefont {Zhou},
  \citenamefont {Wang}, \citenamefont {Zhang}, \citenamefont {Zhang},
  \citenamefont {Yu},\ and\ \citenamefont {Guo}}]{PhysRevApplied.12.054013}%
  \BibitemOpen
  \bibfield  {author} {\bibinfo {author} {\bibfnamefont {C.}~\bibnamefont
  {Zhou}}, \bibinfo {author} {\bibfnamefont {X.}~\bibnamefont {Wang}}, \bibinfo
  {author} {\bibfnamefont {Y.}~\bibnamefont {Zhang}}, \bibinfo {author}
  {\bibfnamefont {Z.}~\bibnamefont {Zhang}}, \bibinfo {author} {\bibfnamefont
  {S.}~\bibnamefont {Yu}},\ and\ \bibinfo {author} {\bibfnamefont
  {H.}~\bibnamefont {Guo}},\ }\bibfield  {title} {\bibinfo {title}
  {Continuous-variable quantum key distribution with rateless reconciliation
  protocol},\ }\href {https://doi.org/10.1103/PhysRevApplied.12.054013}
  {\bibfield  {journal} {\bibinfo  {journal} {Phys. Rev. Applied}\ }\textbf
  {\bibinfo {volume} {12}},\ \bibinfo {pages} {054013} (\bibinfo {year}
  {2019})}\BibitemShut {NoStop}%
\bibitem [{\citenamefont {Leverrier}\ and\ \citenamefont
  {Grangier}(2009)}]{leverrier2009unconditional}%
  \BibitemOpen
  \bibfield  {author} {\bibinfo {author} {\bibfnamefont {A.}~\bibnamefont
  {Leverrier}}\ and\ \bibinfo {author} {\bibfnamefont {P.}~\bibnamefont
  {Grangier}},\ }\bibfield  {title} {\bibinfo {title} {Unconditional security
  proof of long-distance continuous-variable quantum key distribution with
  discrete modulation},\ }\href
  {https://doi.org/10.1103/PhysRevLett.102.180504} {\bibfield  {journal}
  {\bibinfo  {journal} {Phys. Rev. Lett.}\ }\textbf {\bibinfo {volume} {102}},\
  \bibinfo {pages} {180504} (\bibinfo {year} {2009})}\BibitemShut {NoStop}%
\bibitem [{\citenamefont {Xuan}\ \emph {et~al.}(2009)\citenamefont {Xuan},
  \citenamefont {Zhang},\ and\ \citenamefont {Voss}}]{xuan200924}%
  \BibitemOpen
  \bibfield  {author} {\bibinfo {author} {\bibfnamefont {Q.~D.}\ \bibnamefont
  {Xuan}}, \bibinfo {author} {\bibfnamefont {Z.}~\bibnamefont {Zhang}},\ and\
  \bibinfo {author} {\bibfnamefont {P.~L.}\ \bibnamefont {Voss}},\ }\bibfield
  {title} {\bibinfo {title} {A 24 km fiber-based discretely signaled continuous variable quantum key distribution system},\ }\href@noop {} {\bibfield
  {journal} {\bibinfo  {journal} {Opt. Express}\ }\textbf {\bibinfo {volume}
  {17}},\ \bibinfo {pages} {24244} (\bibinfo {year} {2009})}\BibitemShut
  {NoStop}%
\bibitem [{\citenamefont {Hirano}\ \emph {et~al.}(2017)\citenamefont {Hirano},
  \citenamefont {Ichikawa}, \citenamefont {Matsubara}, \citenamefont {Ono},
  \citenamefont {Oguri}, \citenamefont {Namiki}, \citenamefont {Kasai},
  \citenamefont {Matsumoto},\ and\ \citenamefont
  {Tsurumaru}}]{hirano2017implementation}%
  \BibitemOpen
  \bibfield  {author} {\bibinfo {author} {\bibfnamefont {T.}~\bibnamefont
  {Hirano}}, \bibinfo {author} {\bibfnamefont {T.}~\bibnamefont {Ichikawa}},
  \bibinfo {author} {\bibfnamefont {T.}~\bibnamefont {Matsubara}}, \bibinfo
  {author} {\bibfnamefont {M.}~\bibnamefont {Ono}}, \bibinfo {author}
  {\bibfnamefont {Y.}~\bibnamefont {Oguri}}, \bibinfo {author} {\bibfnamefont
  {R.}~\bibnamefont {Namiki}}, \bibinfo {author} {\bibfnamefont
  {K.}~\bibnamefont {Kasai}}, \bibinfo {author} {\bibfnamefont
  {R.}~\bibnamefont {Matsumoto}},\ and\ \bibinfo {author} {\bibfnamefont
  {T.}~\bibnamefont {Tsurumaru}},\ }\bibfield  {title} {\bibinfo {title}
  {Implementation of continuous-variable quantum key distribution with discrete modulation},\ }\href@noop {} {\bibfield  {journal} {\bibinfo  {journal}
  {Quantum Sci. Technol.}\ }\textbf {\bibinfo {volume} {2}},\ \bibinfo {pages}
  {024010} (\bibinfo {year} {2017})}\BibitemShut {NoStop}%
\bibitem [{\citenamefont {Lin}\ and\ \citenamefont
  {L\"utkenhaus}(2020)}]{lin2020trusted}%
  \BibitemOpen
  \bibfield  {author} {\bibinfo {author} {\bibfnamefont {J.}~\bibnamefont
  {Lin}}\ and\ \bibinfo {author} {\bibfnamefont {N.}~\bibnamefont
  {L\"utkenhaus}},\ }\bibfield  {title} {\bibinfo {title} {Trusted detector
  noise analysis for discrete modulation schemes of continuous-variable quantum
  key distribution},\ }\href {https://doi.org/10.1103/PhysRevApplied.14.064030}
  {\bibfield  {journal} {\bibinfo  {journal} {Phys. Rev. Applied}\ }\textbf
  {\bibinfo {volume} {14}},\ \bibinfo {pages} {064030} (\bibinfo {year}
  {2020})}\BibitemShut {NoStop}%
\bibitem [{\citenamefont {Ghalaii}\ \emph {et~al.}(2020)\citenamefont
  {Ghalaii}, \citenamefont {Ottaviani}, \citenamefont {Kumar}, \citenamefont
  {Pirandola},\ and\ \citenamefont {Razavi}}]{ghalaii2020discrete}%
  \BibitemOpen
  \bibfield  {author} {\bibinfo {author} {\bibfnamefont {M.}~\bibnamefont
  {Ghalaii}}, \bibinfo {author} {\bibfnamefont {C.}~\bibnamefont {Ottaviani}},
  \bibinfo {author} {\bibfnamefont {R.}~\bibnamefont {Kumar}}, \bibinfo
  {author} {\bibfnamefont {S.}~\bibnamefont {Pirandola}},\ and\ \bibinfo
  {author} {\bibfnamefont {M.}~\bibnamefont {Razavi}},\ }\bibfield  {title}
  {\bibinfo {title} {Discrete-modulation continuous-variable quantum key
  distribution enhanced by quantum scissors},\ }\href@noop {} {\bibfield
  {journal} {\bibinfo  {journal} {IEEE J. Sel. Areas Commun.}\ }\textbf
  {\bibinfo {volume} {38}},\ \bibinfo {pages} {506} (\bibinfo {year}
  {2020})}\BibitemShut {NoStop}%
\bibitem [{\citenamefont {Ghorai}\ \emph {et~al.}(2019)\citenamefont {Ghorai},
  \citenamefont {Grangier}, \citenamefont {Diamanti},\ and\ \citenamefont
  {Leverrier}}]{ghorai2019asymptotic}%
  \BibitemOpen
  \bibfield  {author} {\bibinfo {author} {\bibfnamefont {S.}~\bibnamefont
  {Ghorai}}, \bibinfo {author} {\bibfnamefont {P.}~\bibnamefont {Grangier}},
  \bibinfo {author} {\bibfnamefont {E.}~\bibnamefont {Diamanti}},\ and\
  \bibinfo {author} {\bibfnamefont {A.}~\bibnamefont {Leverrier}},\ }\bibfield
  {title} {\bibinfo {title} {Asymptotic security of continuous-variable quantum
  key distribution with a discrete modulation},\ }\href
  {https://doi.org/10.1103/PhysRevX.9.021059} {\bibfield  {journal} {\bibinfo
  {journal} {Phys. Rev. X}\ }\textbf {\bibinfo {volume} {9}},\ \bibinfo {pages}
  {021059} (\bibinfo {year} {2019})}\BibitemShut {NoStop}%
\bibitem [{\citenamefont {Lin}\ \emph {et~al.}(2019)\citenamefont {Lin},
  \citenamefont {Upadhyaya},\ and\ \citenamefont
  {L{\"u}tkenhaus}}]{lin2019asymptotic}%
  \BibitemOpen
  \bibfield  {author} {\bibinfo {author} {\bibfnamefont {J.}~\bibnamefont
  {Lin}}, \bibinfo {author} {\bibfnamefont {T.}~\bibnamefont {Upadhyaya}},\
  and\ \bibinfo {author} {\bibfnamefont {N.}~\bibnamefont {L{\"u}tkenhaus}},\
  }\bibfield  {title} {\bibinfo {title} {Asymptotic security analysis of
  discrete-modulated continuous-variable quantum key distribution},\
  }\href@noop {} {\bibfield  {journal} {\bibinfo  {journal} {Phys. Rev. X}\
  }\textbf {\bibinfo {volume} {9}},\ \bibinfo {pages} {041064} (\bibinfo {year}
  {2019})}\BibitemShut {NoStop}%
\bibitem [{\citenamefont {Hu}\ \emph {et~al.}(2021)\citenamefont {Hu},
  \citenamefont {Im}, \citenamefont {Lin}, \citenamefont {L{\"u}tkenhaus},\
  and\ \citenamefont {Wolkowicz}}]{hu2021robust}%
  \BibitemOpen
  \bibfield  {author} {\bibinfo {author} {\bibfnamefont {H.}~\bibnamefont
  {Hu}}, \bibinfo {author} {\bibfnamefont {J.}~\bibnamefont {Im}}, \bibinfo
  {author} {\bibfnamefont {J.}~\bibnamefont {Lin}}, \bibinfo {author}
  {\bibfnamefont {N.}~\bibnamefont {L{\"u}tkenhaus}},\ and\ \bibinfo {author}
  {\bibfnamefont {H.}~\bibnamefont {Wolkowicz}},\ }\href@noop {} {\bibinfo
  {title} {Robust interior point method for quantum key distribution rate
  computation}} (\bibinfo {year} {2021}),\ \Eprint
  {https://arxiv.org/abs/2104.03847} {arXiv:2104.03847} \BibitemShut {NoStop}%
\bibitem [{\citenamefont {Kaur}\ \emph {et~al.}(2021)\citenamefont {Kaur},
  \citenamefont {Guha},\ and\ \citenamefont {Wilde}}]{PhysRevA.103.012412}%
  \BibitemOpen
  \bibfield  {author} {\bibinfo {author} {\bibfnamefont {E.}~\bibnamefont
  {Kaur}}, \bibinfo {author} {\bibfnamefont {S.}~\bibnamefont {Guha}},\ and\
  \bibinfo {author} {\bibfnamefont {M.~M.}\ \bibnamefont {Wilde}},\ }\bibfield
  {title} {\bibinfo {title} {Asymptotic security of discrete-modulation
  protocols for continuous-variable quantum key distribution},\ }\href
  {https://doi.org/10.1103/PhysRevA.103.012412} {\bibfield  {journal} {\bibinfo
   {journal} {Phys. Rev. A}\ }\textbf {\bibinfo {volume} {103}},\ \bibinfo
  {pages} {012412} (\bibinfo {year} {2021})}\BibitemShut {NoStop}%
\bibitem [{\citenamefont {Denys}\ \emph {et~al.}(2021)\citenamefont {Denys},
  \citenamefont {Brown},\ and\ \citenamefont {Leverrier}}]{denys2021explicit}%
  \BibitemOpen
  \bibfield  {author} {\bibinfo {author} {\bibfnamefont {A.}~\bibnamefont
  {Denys}}, \bibinfo {author} {\bibfnamefont {P.}~\bibnamefont {Brown}},\ and\
  \bibinfo {author} {\bibfnamefont {A.}~\bibnamefont {Leverrier}},\ }\bibfield
  {title} {\bibinfo {title} {Explicit asymptotic secret key rate of continuous-variable quantum key distribution with an arbitrary modulation},\ }\href
  {https://doi.org/10.22331/q-2021-09-13-540} {\bibfield  {journal} {\bibinfo
   {journal} {Quantum}\ }\textbf {\bibinfo {volume} {5}},\ \bibinfo
  {pages} {540} (\bibinfo {year} {2021})}\BibitemShut {NoStop}%
\bibitem [{\citenamefont {Matsuura}\ \emph {et~al.}(2021)\citenamefont
  {Matsuura}, \citenamefont {Maeda}, \citenamefont {Sasaki},\ and\
  \citenamefont {Koashi}}]{matsuura2021finite}%
  \BibitemOpen
  \bibfield  {author} {\bibinfo {author} {\bibfnamefont {T.}~\bibnamefont
  {Matsuura}}, \bibinfo {author} {\bibfnamefont {K.}~\bibnamefont {Maeda}},
  \bibinfo {author} {\bibfnamefont {T.}~\bibnamefont {Sasaki}},\ and\ \bibinfo
  {author} {\bibfnamefont {M.}~\bibnamefont {Koashi}},\ }\bibfield  {title}
  {\bibinfo {title} {Finite-size security of continuous-variable quantum key
  distribution with digital signal processing},\ }\href@noop {} {\bibfield
  {journal} {\bibinfo  {journal} {Nat. Commun.}\ }\textbf {\bibinfo {volume}
  {12}},\ \bibinfo {pages} {252} (\bibinfo {year} {2021})}\BibitemShut
  {NoStop}%
\bibitem [{\citenamefont {Devetak}\ and\ \citenamefont
  {Winter}(2005)}]{devetak2005distillation}%
  \BibitemOpen
  \bibfield  {author} {\bibinfo {author} {\bibfnamefont {I.}~\bibnamefont
  {Devetak}}\ and\ \bibinfo {author} {\bibfnamefont {A.}~\bibnamefont
  {Winter}},\ }\bibfield  {title} {\bibinfo {title} {Distillation of secret key and entanglement from quantum states},\ }\href
  {https://doi.org/10.1098/rspa.2004.1372} {\bibfield  {journal} {\bibinfo
  {journal} {Proc. R. Soc. A.}\ }\textbf {\bibinfo {volume} {461}},\ \bibinfo
  {pages} {207} (\bibinfo {year} {2005})}\BibitemShut {NoStop}%
\bibitem [{\citenamefont {Coles}\ \emph {et~al.}(2016)\citenamefont {Coles},
  \citenamefont {Metodiev},\ and\ \citenamefont
  {L{\"u}tkenhaus}}]{coles2016numerical}%
  \BibitemOpen
  \bibfield  {author} {\bibinfo {author} {\bibfnamefont {P.~J.}\ \bibnamefont
  {Coles}}, \bibinfo {author} {\bibfnamefont {E.~M.}\ \bibnamefont
  {Metodiev}},\ and\ \bibinfo {author} {\bibfnamefont {N.}~\bibnamefont
  {L{\"u}tkenhaus}},\ }\bibfield  {title} {\bibinfo {title} {Numerical approach for unstructured quantum key distribution},\ }\href@noop {} {\bibfield
  {journal} {\bibinfo  {journal} {Nat. Commun.}\ }\textbf {\bibinfo {volume}
  {7}},\ \bibinfo {pages} {11712} (\bibinfo {year} {2016})}\BibitemShut
  {NoStop}%
\bibitem [{\citenamefont {Winick}\ \emph {et~al.}(2018)\citenamefont {Winick},
  \citenamefont {L{\"u}tkenhaus},\ and\ \citenamefont
  {Coles}}]{winick2018reliable}%
  \BibitemOpen
  \bibfield  {author} {\bibinfo {author} {\bibfnamefont {A.}~\bibnamefont
  {Winick}}, \bibinfo {author} {\bibfnamefont {N.}~\bibnamefont
  {L{\"u}tkenhaus}},\ and\ \bibinfo {author} {\bibfnamefont {P.~J.}\
  \bibnamefont {Coles}},\ }\bibfield  {title} {\bibinfo {title} {Reliable
  numerical key rates for quantum key distribution},\ }\href@noop {} {\bibfield
   {journal} {\bibinfo  {journal} {Quantum}\ }\textbf {\bibinfo {volume} {2}},\
  \bibinfo {pages} {77} (\bibinfo {year} {2018})}\BibitemShut {NoStop}%
\bibitem [{\citenamefont {Kanitschar}\ and\ \citenamefont
  {Pacher}(2021)}]{kanitschar2021postselection}%
  \BibitemOpen
  \bibfield  {author} {\bibinfo {author} {\bibfnamefont {F.}~\bibnamefont
  {Kanitschar}}\ and\ \bibinfo {author} {\bibfnamefont {C.}~\bibnamefont
  {Pacher}},\ }\href@noop {} {\bibinfo {title} {Postselection strategies for
  continuous-variable quantum key distribution protocols with quadrature
  phase-shift keying modulation}} (\bibinfo {year} {2021}),\ \Eprint
  {https://arxiv.org/abs/2104.09454} {arXiv:2104.09454} \BibitemShut
  {NoStop}%
\bibitem [{\citenamefont {Zhou}\ \emph {et~al.}(2021)\citenamefont {Zhou},
  \citenamefont {Liu}, \citenamefont {Liu}, \citenamefont {Li}, \citenamefont
  {Bai}, \citenamefont {Xue}, \citenamefont {Fu}, \citenamefont {Yin},\ and\
  \citenamefont {Chen}}]{zhou2021machine}%
  \BibitemOpen
  \bibfield  {author} {\bibinfo {author} {\bibfnamefont {M.-G.}\ \bibnamefont
  {Zhou}}, \bibinfo {author} {\bibfnamefont {Z.-P.}\ \bibnamefont {Liu}},
  \bibinfo {author} {\bibfnamefont {W.-B.}\ \bibnamefont {Liu}}, \bibinfo
  {author} {\bibfnamefont {C.-L.}\ \bibnamefont {Li}}, \bibinfo {author}
  {\bibfnamefont {J.-L.}\ \bibnamefont {Bai}}, \bibinfo {author} {\bibfnamefont
  {Y.-R.}\ \bibnamefont {Xue}}, \bibinfo {author} {\bibfnamefont
  {Y.}~\bibnamefont {Fu}}, \bibinfo {author} {\bibfnamefont {H.-L.}\
  \bibnamefont {Yin}},\ and\ \bibinfo {author} {\bibfnamefont {Z.-B.}\
  \bibnamefont {Chen}},\ }\href@noop {} {\bibinfo {title} {Machine learning for
  secure key rate in continuous-variable quantum key distribution}} (\bibinfo
  {year} {2021}),\ \Eprint {https://arxiv.org/abs/2108.02578}
  {arXiv:2108.02578} \BibitemShut {NoStop}%
\bibitem [{\citenamefont {Zhao}\ \emph {et~al.}(2009)\citenamefont {Zhao},
  \citenamefont {Heid}, \citenamefont {Rigas},\ and\ \citenamefont
  {L\"utkenhaus}}]{PhysRevA.79.012307}%
  \BibitemOpen
  \bibfield  {author} {\bibinfo {author} {\bibfnamefont {Y.-B.}\ \bibnamefont
  {Zhao}}, \bibinfo {author} {\bibfnamefont {M.}~\bibnamefont {Heid}}, \bibinfo
  {author} {\bibfnamefont {J.}~\bibnamefont {Rigas}},\ and\ \bibinfo {author}
  {\bibfnamefont {N.}~\bibnamefont {L\"utkenhaus}},\ }\bibfield  {title}
  {\bibinfo {title} {Asymptotic security of binary modulated
  continuous-variable quantum key distribution under collective attacks},\
  }\href@noop {} {\bibfield  {journal} {\bibinfo  {journal} {Phys. Rev. A}\
  }\textbf {\bibinfo {volume} {79}},\ \bibinfo {pages} {012307} (\bibinfo
  {year} {2009})}\BibitemShut {NoStop}%
\bibitem [{\citenamefont {Br\'adler}\ and\ \citenamefont
  {Weedbrook}(2018)}]{PhysRevA.97.022310}%
  \BibitemOpen
  \bibfield  {author} {\bibinfo {author} {\bibfnamefont {K.}~\bibnamefont
  {Br\'adler}}\ and\ \bibinfo {author} {\bibfnamefont {C.}~\bibnamefont
  {Weedbrook}},\ }\bibfield  {title} {\bibinfo {title} {Security proof of
  continuous-variable quantum key distribution using three coherent states},\
  }\href@noop {} {\bibfield  {journal} {\bibinfo  {journal} {Phys. Rev. A}\
  }\textbf {\bibinfo {volume} {97}},\ \bibinfo {pages} {022310} (\bibinfo
  {year} {2018})}\BibitemShut {NoStop}%
\bibitem [{\citenamefont {Weedbrook}\ \emph {et~al.}(2004)\citenamefont
  {Weedbrook}, \citenamefont {Lance}, \citenamefont {Bowen}, \citenamefont
  {Symul}, \citenamefont {Ralph},\ and\ \citenamefont
  {Lam}}]{PhysRevLett.93.170504}%
  \BibitemOpen
  \bibfield  {author} {\bibinfo {author} {\bibfnamefont {C.}~\bibnamefont
  {Weedbrook}}, \bibinfo {author} {\bibfnamefont {A.~M.}\ \bibnamefont
  {Lance}}, \bibinfo {author} {\bibfnamefont {W.~P.}\ \bibnamefont {Bowen}},
  \bibinfo {author} {\bibfnamefont {T.}~\bibnamefont {Symul}}, \bibinfo
  {author} {\bibfnamefont {T.~C.}\ \bibnamefont {Ralph}},\ and\ \bibinfo
  {author} {\bibfnamefont {P.~K.}\ \bibnamefont {Lam}},\ }\bibfield  {title}
  {\bibinfo {title} {Quantum cryptography without switching},\ }\href
  {https://doi.org/10.1103/PhysRevLett.93.170504} {\bibfield  {journal}
  {\bibinfo  {journal} {Phys. Rev. Lett.}\ }\textbf {\bibinfo {volume} {93}},\
  \bibinfo {pages} {170504} (\bibinfo {year} {2004})}\BibitemShut {NoStop}%
\bibitem [{\citenamefont {Lance}\ \emph {et~al.}(2005)\citenamefont {Lance},
  \citenamefont {Symul}, \citenamefont {Sharma}, \citenamefont {Weedbrook},
  \citenamefont {Ralph},\ and\ \citenamefont {Lam}}]{lance2005no}%
  \BibitemOpen
  \bibfield  {author} {\bibinfo {author} {\bibfnamefont {A.~M.}\ \bibnamefont
  {Lance}}, \bibinfo {author} {\bibfnamefont {T.}~\bibnamefont {Symul}},
  \bibinfo {author} {\bibfnamefont {V.}~\bibnamefont {Sharma}}, \bibinfo
  {author} {\bibfnamefont {C.}~\bibnamefont {Weedbrook}}, \bibinfo {author}
  {\bibfnamefont {T.~C.}\ \bibnamefont {Ralph}},\ and\ \bibinfo {author}
  {\bibfnamefont {P.~K.}\ \bibnamefont {Lam}},\ }\bibfield  {title} {\bibinfo
  {title} {No-switching quantum key distribution using broadband modulated
  coherent light},\ }\href@noop {} {\bibfield  {journal} {\bibinfo  {journal}
  {Phys. Rev. Lett.}\ }\textbf {\bibinfo {volume} {95}},\ \bibinfo {pages}
  {180503} (\bibinfo {year} {2005})}\BibitemShut {NoStop}%
\bibitem [{\citenamefont {Brunner}\ \emph {et~al.}()\citenamefont {Brunner},
  \citenamefont {Comandar}, \citenamefont {Karinou}, \citenamefont {Bettelli},
  \citenamefont {Hillerkuss}, \citenamefont {Fung}, \citenamefont {Wang},
  \citenamefont {Mikroulis}, \citenamefont {Kuschnerov}, \citenamefont {Poppe}
  \emph {et~al.}}]{brunner2017low}%
  \BibitemOpen
  \bibfield  {author} {\bibinfo {author} {\bibfnamefont {H.~H.}\ \bibnamefont
  {Brunner}}, \bibinfo {author} {\bibfnamefont {L.~C.}\ \bibnamefont
  {Comandar}}, \bibinfo {author} {\bibfnamefont {F.}~\bibnamefont {Karinou}},
  \bibinfo {author} {\bibfnamefont {S.}~\bibnamefont {Bettelli}}, \bibinfo
  {author} {\bibfnamefont {D.}~\bibnamefont {Hillerkuss}}, \bibinfo {author}
  {\bibfnamefont {F.}~\bibnamefont {Fung}}, \bibinfo {author} {\bibfnamefont
  {D.}~\bibnamefont {Wang}}, \bibinfo {author} {\bibfnamefont {S.}~\bibnamefont
  {Mikroulis}}, \bibinfo {author} {\bibfnamefont {M.}~\bibnamefont
  {Kuschnerov}}, \bibinfo {author} {\bibfnamefont {A.}~\bibnamefont {Poppe}},
  \emph {et~al.},\ }\bibfield  {title} {\bibinfo {title} {Low-noise,
  low-complexity cv-qkd architecture},\ }in\ \href@noop {} {\emph {\bibinfo
  {booktitle} {QCrypt 2017(Cambridge)}}}\BibitemShut {NoStop}%
\bibitem [{\citenamefont {Chin}\ \emph {et~al.}(2021)\citenamefont {Chin},
  \citenamefont {Jain}, \citenamefont {Zibar}, \citenamefont {Andersen},\ and\
  \citenamefont {Gehring}}]{chin2021machine}%
  \BibitemOpen
  \bibfield  {author} {\bibinfo {author} {\bibfnamefont {H.-M.}\ \bibnamefont
  {Chin}}, \bibinfo {author} {\bibfnamefont {N.}~\bibnamefont {Jain}}, \bibinfo
  {author} {\bibfnamefont {D.}~\bibnamefont {Zibar}}, \bibinfo {author}
  {\bibfnamefont {U.~L.}\ \bibnamefont {Andersen}},\ and\ \bibinfo {author}
  {\bibfnamefont {T.}~\bibnamefont {Gehring}},\ }\bibfield  {title} {\bibinfo
  {title} {Machine learning aided carrier recovery in continuous-variable
  quantum key distribution},\ }\href@noop {} {\bibfield  {journal} {\bibinfo
  {journal} {npj Quantum Inf.}\ }\textbf {\bibinfo {volume} {7}},\ \bibinfo
  {pages} {20} (\bibinfo {year} {2021})}\BibitemShut {NoStop}%
\bibitem [{\citenamefont {Jain}\ \emph {et~al.}(2021)\citenamefont {Jain},
  \citenamefont {Derkach}, \citenamefont {Chin}, \citenamefont {Filip},
  \citenamefont {Andersen}, \citenamefont {Usenko},\ and\ \citenamefont
  {Gehring}}]{Jain_2021}%
  \BibitemOpen
  \bibfield  {author} {\bibinfo {author} {\bibfnamefont {N.}~\bibnamefont
  {Jain}}, \bibinfo {author} {\bibfnamefont {I.}~\bibnamefont {Derkach}},
  \bibinfo {author} {\bibfnamefont {H.-M.}\ \bibnamefont {Chin}}, \bibinfo
  {author} {\bibfnamefont {R.}~\bibnamefont {Filip}}, \bibinfo {author}
  {\bibfnamefont {U.~L.}\ \bibnamefont {Andersen}}, \bibinfo {author}
  {\bibfnamefont {V.~C.}\ \bibnamefont {Usenko}},\ and\ \bibinfo {author}
  {\bibfnamefont {T.}~\bibnamefont {Gehring}},\ }\bibfield  {title} {\bibinfo
  {title} {Modulation leakage vulnerability in continuous-variable quantum key
  distribution},\ }\href@noop {} {\bibfield  {journal} {\bibinfo  {journal}
  {Quantum Sci. Technol.}\ }\textbf {\bibinfo {volume} {6}},\ \bibinfo {pages}
  {045001} (\bibinfo {year} {2021})}\BibitemShut {NoStop}%
\bibitem [{\citenamefont {Caves}\ and\ \citenamefont
  {Drummond}(1994)}]{RevModPhys.66.481}%
  \BibitemOpen
  \bibfield  {author} {\bibinfo {author} {\bibfnamefont {C.~M.}\ \bibnamefont
  {Caves}}\ and\ \bibinfo {author} {\bibfnamefont {P.~D.}\ \bibnamefont
  {Drummond}},\ }\bibfield  {title} {\bibinfo {title} {Quantum limits on
  bosonic communication rates},\ }\href
  {https://doi.org/10.1103/RevModPhys.66.481} {\bibfield  {journal} {\bibinfo
  {journal} {Rev. Mod. Phys.}\ }\textbf {\bibinfo {volume} {66}},\ \bibinfo
  {pages} {481} (\bibinfo {year} {1994})}\BibitemShut {NoStop}%
\bibitem [{\citenamefont {Leonhardt}(1997)}]{leonhardt1997measuring}%
  \BibitemOpen
  \bibfield  {author} {\bibinfo {author} {\bibfnamefont {U.}~\bibnamefont
  {Leonhardt}},\ }\href@noop {} {\emph {\bibinfo {title} {Measuring the quantum
  state of light}}},\ Vol.~\bibinfo {volume} {22}\ (\bibinfo  {publisher}
  {Cambridge university press},\ \bibinfo {city} {Cambridge, England},\ \bibinfo {year} {1997})\BibitemShut {NoStop}%
\bibitem [{\citenamefont {Dall'Arno}\ \emph {et~al.}(2010)\citenamefont
  {Dall'Arno}, \citenamefont {D'Ariano},\ and\ \citenamefont
  {Sacchi}}]{PhysRevA.82.042315}%
  \BibitemOpen
  \bibfield  {author} {\bibinfo {author} {\bibfnamefont {M.}~\bibnamefont
  {Dall'Arno}}, \bibinfo {author} {\bibfnamefont {G.~M.}\ \bibnamefont
  {D'Ariano}},\ and\ \bibinfo {author} {\bibfnamefont {M.~F.}\ \bibnamefont
  {Sacchi}},\ }\bibfield  {title} {\bibinfo {title} {Purification of noisy
  quantum measurements},\ }\href {https://doi.org/10.1103/PhysRevA.82.042315}
  {\bibfield  {journal} {\bibinfo  {journal} {Phys. Rev. A}\ }\textbf {\bibinfo
  {volume} {82}},\ \bibinfo {pages} {042315} (\bibinfo {year}
  {2010})}\BibitemShut {NoStop}%
\bibitem [{\citenamefont {Laudenbach}\ \emph {et~al.}(2018)\citenamefont
  {Laudenbach}, \citenamefont {Pacher}, \citenamefont {Fung}, \citenamefont
  {Poppe}, \citenamefont {Peev}, \citenamefont {Schrenk}, \citenamefont
  {Hentschel}, \citenamefont {Walther},\ and\ \citenamefont
  {H\"ubel}}]{CVreview}%
  \BibitemOpen
  \bibfield  {author} {\bibinfo {author} {\bibfnamefont {F.}~\bibnamefont
  {Laudenbach}}, \bibinfo {author} {\bibfnamefont {C.}~\bibnamefont {Pacher}},
  \bibinfo {author} {\bibfnamefont {C.-H.~F.}\ \bibnamefont {Fung}}, \bibinfo
  {author} {\bibfnamefont {A.}~\bibnamefont {Poppe}}, \bibinfo {author}
  {\bibfnamefont {M.}~\bibnamefont {Peev}}, \bibinfo {author} {\bibfnamefont
  {B.}~\bibnamefont {Schrenk}}, \bibinfo {author} {\bibfnamefont
  {M.}~\bibnamefont {Hentschel}}, \bibinfo {author} {\bibfnamefont
  {P.}~\bibnamefont {Walther}},\ and\ \bibinfo {author} {\bibfnamefont
  {H.}~\bibnamefont {H\"ubel}},\ }\bibfield  {title} {\bibinfo {title}
  {Continuous-variable quantum key distribution with gaussian modulation - the
  theory of practical implementations},\ }\href@noop {} {\bibfield  {journal}
  {\bibinfo  {journal} {Adv. Quantum Technol.}\ }\textbf {\bibinfo {volume}
  {1}},\ \bibinfo {pages} {1800011} (\bibinfo {year} {2018})}\BibitemShut
  {NoStop}%
\bibitem [{\citenamefont {Dong}\ \emph {et~al.}(2012)\citenamefont {Dong},
  \citenamefont {Chen}, \citenamefont {Xie}, \citenamefont {Buhl},\ and\
  \citenamefont {Chen}}]{po201250gbs}%
  \BibitemOpen
  \bibfield  {author} {\bibinfo {author} {\bibfnamefont {P.}~\bibnamefont
  {Dong}}, \bibinfo {author} {\bibfnamefont {L.}~\bibnamefont {Chen}}, \bibinfo
  {author} {\bibfnamefont {C.}~\bibnamefont {Xie}}, \bibinfo {author}
  {\bibfnamefont {L.~L.}\ \bibnamefont {Buhl}},\ and\ \bibinfo {author}
  {\bibfnamefont {Y.-K.}\ \bibnamefont {Chen}},\ }\bibfield  {title} {\bibinfo
  {title} {50-gb/s silicon quadrature phase-shift keying modulator},\ }\href
  {https://doi.org/10.1364/OE.20.021181} {\bibfield  {journal} {\bibinfo
  {journal} {Opt. Express}\ }\textbf {\bibinfo {volume} {20}},\ \bibinfo
  {pages} {21181} (\bibinfo {year} {2012})}\BibitemShut {NoStop}%
\bibitem [{\citenamefont {Upadhyaya}\ \emph {et~al.}(2021)\citenamefont
  {Upadhyaya}, \citenamefont {van Himbeeck}, \citenamefont {Lin},\ and\
  \citenamefont {L\"utkenhaus}}]{upadhyaya2021dimension}%
  \BibitemOpen
  \bibfield  {author} {\bibinfo {author} {\bibfnamefont {T.}~\bibnamefont
  {Upadhyaya}}, \bibinfo {author} {\bibfnamefont {T.}~\bibnamefont {van
  Himbeeck}}, \bibinfo {author} {\bibfnamefont {J.}~\bibnamefont {Lin}},\ and\
  \bibinfo {author} {\bibfnamefont {N.}~\bibnamefont {L\"utkenhaus}},\
  }\bibfield  {title} {\bibinfo {title} {Dimension reduction in quantum key
  distribution for continuous- and discrete-variable protocols},\ }\href@noop
  {} {\bibfield  {journal} {\bibinfo  {journal} {PRX Quantum}\ }\textbf
  {\bibinfo {volume} {2}},\ \bibinfo {pages} {020325} (\bibinfo {year}
  {2021})}\BibitemShut {NoStop}%
\bibitem [{\citenamefont {Renner}(2008)}]{renner2008security}%
  \BibitemOpen
  \bibfield  {author} {\bibinfo {author} {\bibfnamefont {R.}~\bibnamefont
  {Renner}},\ }\bibfield  {title} {\bibinfo {title} {Security of quantum key
  distribution},\ }\href@noop {} {\bibfield  {journal} {\bibinfo  {journal}
  {Int. J. Quantum Inf.}\ }\textbf {\bibinfo {volume} {6}},\ \bibinfo {pages}
  {1} (\bibinfo {year} {2008})}\BibitemShut {NoStop}%
\bibitem [{\citenamefont {Grant}\ and\ \citenamefont {Boyd}(2014)}]{cvx}%
  \BibitemOpen
  \bibfield  {author} {\bibinfo {author} {\bibfnamefont {M.}~\bibnamefont
  {Grant}}\ and\ \bibinfo {author} {\bibfnamefont {S.}~\bibnamefont {Boyd}},\
  }\href@noop {} {\bibinfo {title} {{CVX}: Matlab software for disciplined
  convex programming, version 2.1}},\ \bibinfo {howpublished}
  {\url{http://cvxr.com/cvx}} (\bibinfo {year} {2014})\BibitemShut {NoStop}%
\bibitem [{\citenamefont {Grant}\ and\ \citenamefont {Boyd}(2008)}]{gb08}%
  \BibitemOpen
  \bibfield  {author} {\bibinfo {author} {\bibfnamefont {M.}~\bibnamefont
  {Grant}}\ and\ \bibinfo {author} {\bibfnamefont {S.}~\bibnamefont {Boyd}},\
  }\bibfield  {title} {\bibinfo {title} {Graph implementations for nonsmooth
  convex programs},\ }in\ \href@noop {} {\emph {\bibinfo {booktitle} {Recent
  Advances in Learning and Control}}},\ \bibinfo {series and number} {Lecture
  Notes in Control and Information Sciences},\ \bibinfo {editor} {edited by\
  \bibinfo {editor} {\bibfnamefont {V.}~\bibnamefont {Blondel}}, \bibinfo
  {editor} {\bibfnamefont {S.}~\bibnamefont {Boyd}},\ and\ \bibinfo {editor}
  {\bibfnamefont {H.}~\bibnamefont {Kimura}}}\ (\bibinfo  {publisher}
  {Springer-Verlag Limited},\ \bibinfo {year} {2008})\ pp.\ \bibinfo {pages}
  {95--110},\ \bibinfo {note}
  {\url{http://stanford.edu/~boyd/graph_dcp.html}}\BibitemShut {NoStop}%
\bibitem [{\citenamefont {Liu}\ and\ \citenamefont {Li}()}]{wherecode}%
  \BibitemOpen
  \bibfield  {author} {\bibinfo {author} {\bibfnamefont {W.-B.}\ \bibnamefont
  {Liu}}\ and\ \bibinfo {author} {\bibfnamefont {C.-L.}\ \bibnamefont {Li}},\
  }\href@noop {} {\bibinfo {title} {Codes of key rate calculation}},\ \bibinfo
  {howpublished}
  {\url{https://github.com/hechuwuliu/HD-QPSK-CVQKD}}\BibitemShut {NoStop}%
\end{thebibliography}
\end{document}